\documentclass{aa}
\usepackage{graphicx}
\usepackage{amsmath}
\usepackage{siunitx}
\usepackage[capitalize]{cleveref}
\usepackage{booktabs}
\usepackage{txfonts}
\usepackage{natbib}
\bibpunct{(}{)}{;}{a}{}{,} 
\usepackage[svgnames]{xcolor}
\usepackage{bm}
\usepackage[normalem]{ulem}

\newcommand{\msun}{\ensuremath{\textit{M}_\odot}}

\newcommand{\ord}[1]{\ensuremath{\mathcal{O}\left({#1}\right)}}
\renewcommand{\rho}{\varrho}

\newcommand{\avg}[1]{\left< #1 \right>}

\renewcommand{\u}{{\vec u}}
\newcommand{\f}{{\vec f}}
\newcommand{\s}{{\vec s}}

\newcommand{\nU}{{\vec U}}
\newcommand{\nF}{{\vec F}}
\newcommand{\nS}{{\vec S}}

\newcommand{\const}{\ensuremath{\text{\textit{\emph{const.}}}}}
\newcommand{\half}{\ensuremath{\frac{1}{2}}}
\newcommand{\oo}[1]{\ensuremath{\frac{1}{#1}}}
\newcommand{\del}{\partial}
\newcommand{\D}[2][ ]{\frac{\partial #1}{\partial #2}}

\newcommand{\DD}[2]{\frac{\partial #1}{\partial #2}}
\newcommand{\Mcut}{{\ensuremath{M_{\rm cut}}}\xspace}
\newcommand{\Mpcut}{{\ensuremath{M_{\rm cut}^p}}\xspace}
\newcommand{\fapdiff}{{\ensuremath{f_a^\text{p}}}\xspace}

\newcommand{\x}{\ensuremath{\vec x}}
\newcommand{\g}{\ensuremath{\vec g}}
\renewcommand{\rho}{\varrho}
\newcommand{\vel}{\vec v}
\newcommand{\brunt}{Brunt--V\"ais\"al\"a\xspace}
\newcommand{\tBV}{\xspace\ensuremath{t_\text{BV}}\xspace}
\newcommand{\tSC}{\xspace\ensuremath{t_\text{SC}}\xspace}

\newcommand{\df}[2]{\frac{\partial #1}{\partial #2}}

\newcommand{\hvF}{\hat{\vec F}}

\newcommand{\alphabeta}{\mbox{$\alpha$-$\beta$}\xspace}
\newcommand{\leroux}{Cargo--LeRoux\xspace}

\newcommand{\wbing}{well-balancing\xspace}
\newcommand{\wbd}{well-balanced\xspace}
\newcommand{\ausmpup}{AUSM${}^+$-up\xspace}
\newcommand{\ausmup}{\ausmpup}
\newcommand{\ausm}{AUSM${}^+$-up\xspace}
\newcommand{\ausmp}{\ensuremath{\text{AUSM}^{+}_\text{B}\text{-up}}\xspace}
\newcommand{\rhoP}{\ensuremath{\rho\text{-}p}\xspace}
\newcommand{\rhoT}{\ensuremath{\rho\text{-}T}\xspace}
\newcommand{\eos}{EoS\xspace}
\newcommand{\slh}{SLH\xspace}

\newcommand{\supad}{\ensuremath{\Delta\nabla}\xspace}
\newcommand{\stab}{\text{SZ}}
\newcommand{\cz}{\text{CZ}}
\newcommand{\ycs}{\ensuremath{y_{\cz,\stab}}\xspace}
\newcommand{\ma}{{\ensuremath{\text{Ma}}}\xspace}
\newcommand{\Mo}{{\ensuremath{M_\mathrm{o}}}\xspace}
\newcommand{\mrms}{\ensuremath{\ma_\text{rms}}\xspace}
\newcommand{\tconv}{\ensuremath{\tau_\text{conv}}\xspace}

\newcommand{\dote}{\ensuremath{\dot{e}_0}\xspace}

\titlerunning{Well-balancing in astrophysical hydrodynamics}
\begin{document}
\title{Well-balanced treatment of gravity in astrophysical fluid dynamics simulations at low Mach numbers}
%
%
\author{
    P.~V.~F.~Edelmann\inst{1,2,3}\and
    L.~Horst\inst{2}\and
    J.~P.~Berberich\inst{4}
    \and R.~Andrassy\inst{2}\and
    J.~Higl\inst{2}\and
    G.~Leidi\inst{2,5}\and
    C.~Klingenberg\inst{4}\and
    F.~K.~R{\"o}pke\inst{2,6}}
\institute{
    X Computational Physics (XCP) Division and Center for Theoretical Astrophysics (CTA),
    Los Alamos National Laboratory,
    Los Alamos,
    NM 87545,
    USA\\
    \email{philipp@slh-code.org}
    \and
    Heidelberger Institut f{\"u}r Theoretische Studien,
    Schloss-Wolfsbrunnenweg 35,
    D-69118 Heidelberg,
    Germany
    \and
    Faculty of Physics and Astronomy,
    W\"urzburg University,
    Am Hubland,
    D-97074 W\"urzburg, Germany
    \and
    Department of Mathematics,
    W\"urzburg University,
    Emil-Fischer-Str. 40,
    D-97074 W\"urzburg, Germany
    \and
    Zentrum f\"ur Astronomie der Universit\"at Heidelberg,
    Astronomisches Rechen-Institut,
    M\"onchhofstr. 12-14,
    D-69120 Heidelberg, Germany
    \and
    Zentrum f\"ur Astronomie der Universit\"at Heidelberg,
    Institut f\"ur Theoretische Astrophysik,
    Philosophenweg 12,
    D-69120 Heidelberg, Germany
}

\date{Received xxxx xx, xxxx / accepted xxxx xx, xxxx}
%
%
\abstract
%
{Accurate simulations of flows in stellar interiors are crucial to improving
our understanding of stellar structure and evolution. Because the typically
slow flows are merely tiny perturbations on top of a close balance between
gravity and the pressure gradient, such simulations place heavy demands on
numerical hydrodynamics schemes.}
{We demonstrate how discretization errors on grids of reasonable size can
lead to spurious flows orders of magnitude faster than the physical flow.
Well-balanced numerical schemes can deal with this problem.}
{Three such schemes were applied in the implicit, finite-volume
\textsc{Seven-League Hydro} (\slh) code in combination with a low-Mach-number
numerical flux function. We compare how the schemes perform in four numerical
experiments
addressing some of the challenges imposed by typical problems in stellar
hydrodynamics.}
{We find that the $\alpha$-$\beta$ and deviation well-balancing methods can
accurately maintain hydrostatic solutions provided that gravitational
potential energy is included in the total energy balance. They accurately conserve
minuscule entropy fluctuations advected in an isentropic stratification,
which enables the methods to reproduce the expected scaling of convective flow
speed with the heating rate. The deviation method also substantially
increases accuracy of maintaining stationary orbital motions in a Keplerian
disk on long timescales. The \leroux method fares substantially worse in our
tests, although its simplicity may still offer some merits in certain
situations.}
{Overall, we find the well-balanced treatment of gravity in combination with
low Mach number flux functions essential to reproducing correct physical
solutions to challenging stellar slow-flow problems on affordable collocated
grids.}
\keywords{hydrodynamics -- methods: numerical -- convection}

\maketitle
%

\section{Introduction}
\label{sec:introduction}
Astrophysical modeling often involves self-gravitating fluids. They are
commonly described by the equations of fluid dynamics with a gravitational
source term -- viscous effects are negligible in most astrophysical systems and
therefore the nonviscous Euler equations are used.
Such systems can attain stationary equilibrium configurations in which a
pressure gradient balances gravity, that is hydrostatic equilibrium.
A prominent example are stars, modeled in classical approaches as
spherically symmetric gaseous objects. Apart from this dimensional reduction,
the assumption of hydrostatic equilibrium considerably simplifies the modeling
of the -- in reality rather complex -- structure of stars. The resulting equations of
stellar structure \citep[e.g.,][]{kippenhahn2012a} enable successful
qualitative modeling of the evolution of stars through different stages. The
price for this success is a parametrization of multidimensional and dynamical
processes that limits the predictive power of such theoretical models and
requires their calibration with observations. Recent attempts to simulate
inherently multidimensional and dynamical processes, such as convection in
stellar interiors
\citep[e.g.,][]{browning2004a,meakin2006a,meakin2007a,woodward2015a,rogers2013a,viallet2013b,pratt2016a,mueller2016a,cristini2017a,edelmann2019a,horst2020a},
have tried to overcome this shortcoming.

Such simulations pose a number of challenges to the underlying numerical
techniques. Not only is the range of relevant spatial and temporal scales
excessive, but the flows of interest arise in a configuration that is
often close to hydrostatic equilibrium. This has two implications:
(i) The schemes must be able to preserve hydrostatic equilibrium in stable
setups over a long period of time compared to the typical timescales of the flows
of interest. (ii) The flow speed $v$ expected to arise from a small
perturbation of the equilibrium configuration should be slow compared to the
speed of sound $c$, thus the corresponding Mach number, $\mathcal{M} \equiv
v/c$, is expected to be low.

If we decide to avoid approximating the equations and include all effects of
compressibility, aspect (ii) above calls for special low-Mach-number solvers in
numerical fluid dynamics combined with time-implicit discretization to enable
time steps determined from the actual fluid velocity instead of the speed of
sound as required by the CFL stability criterion \citep{courant1928a} of
time-explicit schemes. The propagation of sound waves is irrelevant for the
problems at hand. Several suitable methods are implemented in the
\textsc{Seven-League Hydro} (\slh) code. Numerical and theoretical details are
discussed in
\citet{barsukow2017b}, \citet{barsukow2017a}, \citet{edelmann2016b}, \citet{miczek2015a}, \citet{edelmann2014a}, and \citet{miczek2013a},
and examples of the application to astrophysical problems can be found in
\citet{horst2020a}, \citet{roepke2018b}, \citet{edelmann2017a}, \citet{michel_phd}, and \citet{bolanos_phd}.

Aspect (i), however, also requires attention.  The condition for hydrostatic
equilibrium is part of the equations of stellar structure, that are discretized
and numerically solved in classical stellar evolution modeling approaches. In
contrast, hydrostatic equilibrium is only a special solution to the full
gravo-hydrodynamic system at the level of the partial differential equations,
but it is not guaranteed that discretizations of these equations can reproduce
the physically correct equilibrium state. This is in particular the case because
gravitational source terms are usually treated in an operator-splitting
approach, resulting in different discretizations of the pressure and gravity
terms. Astrophysical fluid dynamics simulations often employ finite-volume
schemes, in which hydrodynamical flows are modeled with a Godunov-type flux
across cell interfaces. Hydrodynamical quantities are therefore determined at
these locations. The gravitational source term, in contrast, is discretized in a
completely different and independent way. In a second-order code, for example,
it is often calculated using cell-averaged densities assigned to cell centers.
In general, this procedure does not lead to an exact cancellation of gravity and
pressure gradient in hydrostatic configurations. Spurious motions are introduced
that mask the delicate low-Mach-number flows arising from perturbations of this
equilibrium, such as, for instance, convection driven by nuclear energy release.

To overcome the problem of aspect (i), so-called well-balancing methods have
been introduced, which are numerical methods that ensure exact preservation of
certain stationary states.  Methods of this type have predominantly been
developed for the simulation of shallow-water-type models in order to resolve
stationary solutions such as the lake-at-rest solution without numerical
artifacts
\citep[e.g.,][]{brufau2002a,audusse2004a,bermudez1994a,leveque1998b,desveaux2016b,touma2015a,castro2018a,barsukow2020b}.
These stationary states can be described using an algebraic relation, which
favors the development of well-balanced methods. In the simulation of
hydrodynamics under the influence of a gravitational field, the situation is
different, since hydrostatic solutions are described by a differential equation
that admits a large variety of solutions that depend on temperature and chemical
composition profiles, as well as the equation of state (\eos). In practice, the
concrete hydrostatic profile is determined by equations describing physical
processes other than hydrodynamics and gravity, such as thermal and chemical
transport and the change in energy and species abundance due to reactions.

Different approaches can be used to deal with this: The majority of
well-balanced methods for the Euler equations with gravity, for example
\citet{chandrashekar2015a}, \citet{desveaux2016a}, \citet{touma2016a}, and references therein, are
designed to only balance certain classes of hydrostatic states, often
isothermal, polytropic, or isentropic stratifications, under the assumption of
an ideal gas \eos. However, for many astrophysical
applications, in particular, cases involving late stellar evolutionary stages
and massive stars, nonideal effects of the gas may be important. In stellar
interiors, the most important additions to the ideal gas \eos are radiation
pressure and electron degeneracy effects. This requires a more complex -- often
in parts tabulated -- \eos to properly describe the thermodynamical properties
of the gas. We discuss an example of such an \eos in \cref{sec:helmholtz-eos}.
Well-balanced methods which are capable of balancing hydrostatic states for
general \eos have been introduced by
\citet{cargo1994a}, \citet{kaeppeli2014a}, \citet{kaeppeli2016a}, \citet{grosheintz2019a}, \citet{berberich2018a}, \citet{berberich2019a}, \citet{berberich2020c}, \citet{berberich2021a}, and \citet{berberich2021b}.

Most methods that have been discussed in the astrophysical context and
literature
\citep[e.g.,][]{zingale2002a,perego2016a,kaeppeli2011a,kaeppeli2016a,popov2019a}
balance a second-order approximation of the hydrostatic state rather than the
hydrostatic state itself. Another recent approach is the \wbd, all-Mach-number
scheme by \citet{padioleau2019a}. None of these publications tested a
low-Mach-number, well-balanced method in more than one spatial dimension in a
stable stratification over long timescales. As we show in this paper,
long-term stability cannot be automatically inferred from one-dimensional (1D)
tests, yet it is of fundamental importance for applications in stellar
astrophysics.

Using a staggered grid, which in this context means storing pressure on the
cell interfaces instead of the cell centers, can alleviate some of the problems
of \wbing the atmosphere, as shown, for example, in the MUSIC code
\citep[][Sec.~6]{goffrey2017a}. For this approach, it still has to be shown that
convective velocities scale correctly with the strength of the driving force at
low Mach numbers, which we found very challenging in our approach, see
\cref{subsec:convection}.

The methods introduced in \cite{berberich2018a,berberich2019a,berberich2021a}
can balance any hydrostatic stratification exactly. The only assumption is that
the hydrostatic solution to be balanced is known a priori. This poses no severe
restriction for many astrophysical applications where the initial condition is
often constructed under the assumption of hydrostatic equilibrium. An example
are simulations of stellar convection, where the initial model is commonly
derived from classical stellar evolution calculations that by construction
impose hydrostatic equilibrium. In this context \emph{exact \wbing} refers to
preserving an initial state, which can be calculated to arbitrary precision,
and not to the exactness of other input physics, such as the \eos.

Here, we discuss three possible \wbing methods that follow rather
different approaches. The first method extends the work of \citet{cargo1994a}
which only applied to 1D setups into the three-dimensional (3D) case and
achieves \wbing by modifying the pressure part of a general \eos. We refer to
this as the \textit{\leroux} (CL) \wbing method. The other two methods modify
how variables are extrapolated to the cell interfaces. We refer to them as the
\textit{\alphabeta} \wbing \citep{berberich2018a,berberich2019a} and the
\textit{deviation} \wbing method \citep{berberich2021a}.  For these three
schemes, we describe their theoretical background and study their impact on the
accuracy of solutions to a set of simplified test problems, which are designed
to resemble typical situations in astrophysics.

The structure of the paper is as follows: \Cref{sec:equations}
reviews the basic set of equations of fluid dynamics and their implications.
It also introduces the notation that is used in the subsequent sections. In
\cref{sec:discretization} we discuss the discretization of these
equations and describe the \ausmup flux used in the later tests. The
well-balancing schemes are introduced in
\cref{sec:clr,sec:alphabeta,sec:deviation}. In \cref{sec:tests} we test
the applicability of the well-balancing methods and their performance in an
extensive suite of simple application examples. Conclusions are drawn in \cref{sec:conclusions}.

\section{Equations of compressible, ideal hydrodynamics}
\label{sec:equations}
This section introduces the general set of equations that are solved with
the \slh code in their formulation in general coordinates. The following
subsections closely follow the presentation of \citet{miczek2013a}.

\subsection{Compressible Euler equations}
We employ \textit{curvilinear coordinates} $\vec x=(x,y,z)=(x^1,x^2,x^3)$ with
a smooth mapping,
\begin{equation}
\x:\mathbb{R}^3\rightarrow\mathbb{R}^3,\;
\vec\x\mapsto\x(\vec\xi),
\label{def:curvilinear}
\end{equation}
to Cartesian coordinates $\bm\xi=(\xi,\eta,\zeta)=(\xi^1,\xi^2,\xi^3)$. The reasoning here is that the coordinates~$\vec\xi$ simplify the computations, while the coordinates~$\x$ are adapted to the physical object, such as a spherical star.

The compressible Euler equations on curvilinear coordinates then read
\begin{equation}
\label{def:euler3d}
J\frac{\del \u}{\del t} + A_{\xi}\frac{\del\f_{\xi}}{\del \xi} + A_{\eta}\frac{\del\f_{\eta}}{\del \eta} + A_{\zeta}\frac{\del\f_{\zeta}}{\del \zeta} = J \s ,
\end{equation}
with the vector $\u$ of conserved variables and the fluxes $\f_{\xi^l}$ given by
\begin{equation}
\u = \begin{pmatrix}
\rho\\\rho u\\\rho v\\\rho w\\ E
\end{pmatrix},\quad
\label{def:fluxVectorTransformed}
\f_{\xi^l } =
\begin{pmatrix}
\rho \, \vec n_{\xi^l}^T \vel \\
\rho u \,\vec n_{\xi^l}^T \vel + (n_{\xi^l})_{x} p\\ 
\rho v \,\vec n_{\xi^l}^T \vel + (n_{\xi^l})_{y} p\\ 
\rho w \,\vec n_{\xi^l}^T \vel + (n_{\xi^l})_{z} p\\ 
\vec n_{\xi^i}^T \vel \left( E + p \right)
\end{pmatrix},
\end{equation}
for $l=1,2,3$. Here, density and pressure are denoted by $\rho$ and $p$,
respectively. The velocity vector, expressed through its curvilinear components, reads
$\vel=(u,v,w)$ and enters the equation for the total energy density
$E=\rho\epsilon+\half \rho|\vel|^2 + \rho\phi$ with the specific energy
$\epsilon$ and the gravitational potential $\phi$. The inclusion of the
potential in the total energy does not lead to numerical difficulties here
because it is similar in magnitude to the internal energy in a stellar context
in general and in all the test problems presented in \cref{sec:tests} in
particular. This is possibly different in other situations, where one of the
energies is much larger and cancellation errors can become a problem.

The Euler equations \eqref{def:euler3d} in their curvilinear form in depend on the
derivatives of the coordinate transformation. Its Jacobi determinant is
\begin{equation}
J=\left|\frac{\del \vec x}{\del\vec \xi}\right| = \sum_{l,m,n=1}^3\epsilon_{lmn}\,\DD{x^l}{\xi} \, \DD{x^m}{\eta} \, \DD{x^n}{\zeta},
\label{def:JacobiDet3d}
\end{equation}
where $\epsilon_{lmn}$ is the three-dimensional Levi-Civita symbol.
The normal vector $\vec n_{\xi^l}$ and interface area $A_{\xi^l}$ in $\xi^l$-direction are
\begin{equation}
\label{def:normalVector}
\vec n_{\xi^l} = \frac{J}{A_{\xi^l}}
\begin{pmatrix}
\DD{\xi^l}{x} \\ \DD{\xi^l}{y} \\ \DD{\xi^l}{z}
\end{pmatrix},\quad
A_{\xi^l} = \sqrt{\left(J\,\DD{\xi^l}{x}\right)^2 + \left(J\,\DD{\xi^l}{y}\right)^2 + \left(J\,\DD{\xi^l}{z}\right)^2}.
\end{equation}

External forces that enter \cref{def:euler3d} are -- with an exception
discussed below -- collectively denoted by the source term $\s$. While there
are different possible contributions, for example energy generation due to
nuclear burning, gravity inevitably appears in any astrophysical setup. At the
same time it might pose difficulties
for hydrodynamical codes to maintain hydrostatic solutions to
\cref{def:euler3d} (see \cref{sec:hystat}) if it includes a strong
gravitational source term as is common in the interior of stars. When
gravity is the only source term, the expression for $\s$ reads
\begin{equation}
\label{def:source}
\vec \s = 
\begin{pmatrix}
0\\-\rho\DD{\phi}{x}\\-\rho\DD{\phi}{y}\\-\rho\DD{\phi}{z}\\
0
\end{pmatrix}.
\end{equation}
This adds gravitational forces to the momentum part of \cref{def:euler3d}.
The presence of a gravitational field also affects the evolution of total
energy. We include this effect in the definition of the total energy rather
than the source term, since this treatment significantly improved our accuracy
in numerical experiments. We found this to be crucial in simulations of
low Mach number convection.

The source term adds gravitational force to the momentum equations. For our
current treatment and test setups, the gravitational potential $\phi$ is fixed
in time and changes the mass distribution during the simulation is excluded,
meaning self-gravity is neglected. It is a reasonable simplification if the setup
is very close to hydrostatic equilibrium and the change in the mass
distribution is negligible. Such an approximation, however, fails for stellar
core simulations at later evolutionary stages, where asymmetries in the mass
distribution may arise due to violent convective motions. However, these setups
are not the typical use cases of the \wbing techniques presented in this paper
as deviations from hydrostatic equilibrium are nonnegligible. In our notation
lower indices do not indicate partial derivatives to avoid confusion.

\subsection{Equation of state}
\label{sec:eos}
The common choice to close the Euler system \cref{def:euler3d} is using an EoS.
There are few physically relevant EoS which can be given in a short, explicit
analytical form. Two of these are discussed in the following. We assume that
all components of the gas are in local thermodynamic equilibrium, that is they can
all be described with a common temperature.

\subsubsection{Ideal gas}
\label{sec:ideal_gas}
The ideal gas is one of the simplest EoS, yet with a wide range of
applications. It describes an ensemble of randomly moving, noninteracting
particles in thermodynamic equilibrium. It is an acceptable model for
terrestrial gases, such as air, for which the interactions between the
particles are small. It serves well in the case of a fully ionized plasma, such
as in the interior of stars, as long as the effects of degeneracy and radiation
pressure are small.

The ideal gas pressure is given by
\begin{equation}
p(\rho,\epsilon) = p(\rho,T(\rho,\epsilon)) = \frac{R}{\mu}\rho T(\rho,\epsilon),
\label{eq:ideal_eos}
\end{equation}
with the temperature
\begin{equation}
T(\rho,\epsilon)=\frac{(\gamma-1)\mu}{R}\cdot\frac{\epsilon}{\rho}.
\label{eq:ideal_T}
\end{equation}
The gas constant~$R$ for the ideal gas is
\SI{8.31446261815324e7}{erg.K^{-1}.mol^{-1}}.  The specific heat ratio~$\gamma$
depends on the internal degrees of freedom in the underlying gas mixture,
typical values are $5/3$ for monatomic gases and $7/5$ for diatomic gases.
For our treatment, it is convenient to write the ideal gas EoS in the form of
\cref{eq:ideal_eos} depending on the \emph{temperature} $T$ instead of the
explicit dependence of $\epsilon$. We use this form of the EoS to formulate
hydrostatic equilibria with certain temperature profiles.

\subsubsection{Helmholtz equation of state}
\label{sec:helmholtz-eos}
While the ideal gas is a useful approximation of the EoS of stellar interiors,
it does not capture the effect of partially or fully degenerate electrons or of
radiation pressure. A commonly used EoS that includes these effects to great
precision is the Helmholtz EoS \citep{timmes2000a}. It relies on a
interpolation of the Helmholtz free energy from tabulated values using
biquintic Hermite polynomials. All other quantities are then derived from
expressions involving derivatives of the Helmholtz free energy. This approach
ensures that all thermodynamic consistency relations are fulfilled
automatically. This EoS has a wide range of applicability and serves as a
typical example of a general tabulated EoS, contrasting our approaches to some
well-balanced methods relying on using a specific EoS, such as the ideal gas.

\subsection{Hydrostatic solutions}
\label{sec:hystat}
Except for the very late stages of stellar evolution, stars can be
considered as gaseous spheres, which change only over very long timescales,
much longer that those of the fluid motions. Dynamical processes acting in
the interiors, as for example convective motions, however, evolve on much
shorter timescales. Thus, in hydrodynamical simulations that aim to follow
such fast processes, a star can, to first order, be described as a static
stratification with pressure and density profiles constant in time, that is
\begin{equation}
\vel\equiv \vec 0,\, \rho(t,\x) = \rho(\x),\, \text{and}\, p(t,\x)=p(\x).
\end{equation}
These conditions reduce the first and the last part of \cref{def:euler3d} to
the trivial relations
\begin{equation}
\del_t\rho = 0\qquad \text{and}\qquad \del_t(\rho E)=0.
\end{equation}
The momentum equations lead to the \textit{hydrostatic equation}
\begin{equation}
\nabla p(\rho,T) = -\rho\nabla\phi.
\label{eq:hystat}
\end{equation}
This equation is invariant under transformations between different sets of
curvilinear coordinates. A pair of constant-in-time functions $\rho$ and $p$,
which satisfy \cref{eq:hystat} together with the chosen EoS is called
\emph{hydrostatic solution} or \emph{hydrostatic equilibrium}. Since the \eos
usually depends on temperature, there is in many cases a whole continuum of
hydrostatic solutions rather than uniqueness.

\subsubsection{Convective stability}
\label{sec:brunt}

Depending on the stratification, perturbations to the hydrostatic solution may
lead to dynamical phenomena. One important example is convection, where
hydrostatic equilibrium is not perfectly fulfilled anymore but deviations are
small.

The criterion for stability against convection is typically derived by
considering the behavior of a small fluid element being perturbed from the
surrounding stratification. The frequency at which the element oscillates
around its equilibrium position~$\chi_0$ is called the \brunt frequency~$N$.
Its square is given by
\begin{equation}
N^2 = \left.\D[\phi]{\chi}\oo{\rho_\text{ext}}\left( \D[\rho_\text{int}]{\chi}-\D[\rho_\text{ext}]{\chi} \right)\right|_{\chi_0},
\label{eq:brunt}
\end{equation}
where $\chi$ denotes the vertical coordinate\footnote{that is
the direction opposing the vector of gravitational acceleration},
$\rho_\text{int}$ is the density of the small fluid element, and $\rho_{ext}$
is the density of the background stratification. It is assumed that the fluid
element changes its state adiabatically, that is without exchanging heat with
its surrounding, and the derivative $\partial \rho_\text{int}/ \partial{\chi}$
is interpreted as the adiabatic change of density while maintaining pressure
equilibrium with the background stratification at height~$\chi$. For the full
derivation of \cref{eq:brunt} we refer the reader to any textbook on stellar
astrophysics \citep[e.g.,][]{maeder2009a,kippenhahn2012a}.

It is common to express the gradients in \cref{eq:brunt} in terms of different
variables. In the case of homogeneous composition ($\mu(\x) = \const$)
\cref{eq:brunt} is equivalent to
\begin{equation}
  N^2 = - \D[\phi]{\chi} \D{\chi}\left(\frac{s}{c_p}\right) = - \D[\phi]{\chi} \frac{\delta}{T} \left[ \D[T]{\chi} - \left(\D[T]{\chi}\right)_\text{ad} \right],
  \label{eq:brunt_sT}
\end{equation}
with specific entropy~$s$ and specific heat at constant pressure~$c_p$ and the
equation of state derivative,
\begin{equation}
  \delta = - \D[\ln \rho]{\ln T},
  \label{eq:delta_eos}
\end{equation}
which is 1 in the ideal gas case. The subscript ``ad'' denotes the adiabatic
derivative as mentioned above.

Another common form of this equation is using a variant of the temperature
gradients expressed using pressure as a coordinate,
\begin{equation}
  \nabla = \D[\ln T]{\ln p}.
  \label{eq:nabla}
\end{equation}
Using this definition \cref{eq:brunt} is equivalent to
\begin{equation}
  N^2 = - \D[\phi]{\chi} \frac{\delta}{H_p} \left( \nabla - \nabla_\text{ad} \right),
  \label{eq:brunt_nabla}
\end{equation}
with the pressure scale height,
\begin{equation}
  H_p = - \D[\chi]{\ln p} = - p\D[\chi]{p}.
  \label{eq:Hp}
\end{equation}
We call a hydrostatic equilibrium stable with respect to convection or
convectively stable, if $N^2\geq0$. Otherwise we call it unstable
with respect to convection or convectively unstable.
This is a local definition, which means that a
hydrostatic solution can be convectively stable in one region and convectively
unstable in another. A suitable reference time for convectively stable setups
is the minimal \brunt time
\begin{equation}
  \tBV=\min_{\x \in\Omega} t_\text{BV}^\text{loc}(\x)=\frac{2\pi}{\max_{\x\in\Omega} N(\x)}.
\label{eq:brunt_time}
\end{equation}
It seems to be a natural timescale for the
evolution of small perturbations as explained for example in
\citet{berberich2019a}. For any hydrostatic solution, the value of $N^2$ can
be either calculated analytically (for simple \eos, like the ideal gas \eos)
or numerically for more complex \eos.

Another useful timescale is the sound crossing time~\tSC through the domain.
Similar to \citet{kaeppeli2016a}, we define \tSC in the direction of coordinate
$\xi^l$ as
\begin{equation}
  \tSC = 2 \min_{\xi^i, i\neq l}\int_{\xi^l_L}^{\xi^l_U} d\xi^l \frac{1}{c\left(\xi^1,\xi^2,\xi^3\right)},
\label{eq:sound_crossing}
\end{equation}
with $\xi^l_L$ and $\xi^l_R$ being the lower and upper boundaries of the domain
in that direction. The speed of sound~$c$ is calculated using the equation of
state. An expression for the sound speed using a general equation of state is
given by
\begin{equation}
c(\rho, \epsilon) = \sqrt{\D[p(\rho, \epsilon)]{\rho}+\D[p(\rho, \epsilon)]{\epsilon}\cdot\frac{\epsilon+p(\rho,\epsilon)}{\rho}},
\label{def:speedOfSound}
\end{equation}
where the function for the pressure $p$ comes from an equation of state (see \cref{sec:eos}).

\section{Discretization}
\label{sec:discretization}
Analytic solutions to \cref{def:euler3d} as for example given by the
hydrostatic solution of \cref{sec:hystat} are exceptions and require
special initial conditions. To obtain more general solutions, which also
allow for more complex dynamics such as turbulent convection developed from
perturbations in a hydrostatic stratification, \cref{def:euler3d} needs to be
solved numerically. While there are several different numerical approaches, this
section focuses on the methods that are employed by the \slh code. For a more
general introduction on this topic, see \cite{toro2009a}.

\subsection{Finite-volume scheme}

For the numerical solution, the underlying equations have to be discretized on
a, possibly curvilinear, mesh that resembles the physical spatial domain. This
grid is then mapped to Cartesian coordinates on which the computations are
conducted.  A set of integers $(i,\,j,\,k)$ denotes the center of the
$(i,\,j,\,k)$-th cell while, for example, $(i+1/2,\,j,\,k)$ denotes the
interface between cell $(i,\,j,\,k)$ and $(i+1,\,j,\,k)$. The semi-discrete
finite-volume scheme is obtained by integrating \cref{def:euler3d} over the
cell volume in computational space, leading to
\begin{align}
\label{eq:finiteVolumeScheme}
V_{ijk}\D[\vec U_{ijk}]{t} = -\,
&     A_{i+\half,j,k}(\hat{\vec F}_\xi  )_{i+\half,j,k}
+ A_{i-\half,j,k}(\hat{\vec F}_\xi  )_{i-\half,j,k}\nonumber\\
-\, & A_{i,j+\half,k}(\hat{\vec F}_\eta )_{i,j+\half,k}
+ A_{i,j-\half,k}(\hat{\vec F}_\eta )_{i,j-\half,k} \nonumber\\
-\, & A_{i,j,k+\half}(\hat{\vec F}_\zeta)_{i,j,k+\half}
+ A_{i,j,k-\half}(\hat{\vec F}_\zeta)_{i,j,k-\half}
+V_{ijk}\hat{\vec S}_{ijk},
\end{align}
where $V_{ijk}$ is the volume of the corresponding cell in physical space and
$A_{i+\half,j,k}$, $A_{i,j+\half,k}$, and $A_{i,j,k+\half}$ are the interface
areas of the interfaces in $\xi$, $\eta$, and $\zeta$-direction respectively.
Details on the computation of cell volumes and interface areas are computed to
second order following \cite{kifonidis2012a}. The cell-averaged source term is
approximated to second order by
\begin{equation}
\label{eq:sourceterm_1d}
\hat{\vec S}_{ijk} =
	\rho_{ijk}
\begin{pmatrix}
0\\
\left(g_{x}\right)_{ijk}\\
(g_{y})_{ijk}\\
\left(g_{z}\right)_{ijk}\\
0
\end{pmatrix},
\end{equation}
where $\rho_{ijk}$ is the cell-averaged value of density and the cell-centered
gravitational acceleration
$(g_\chi)_{ijk}=-\left.\DD{\phi}{\chi}\right|_{\x_{ijk}}$ is computed
analytically from the given gravitational potential $\phi$.

In \cref{eq:finiteVolumeScheme}, $\hat{\nF}_{\xi^l}$ is an approximation
of the interface flux for $l=1,2,3$.  There is some freedom in
constructing the approximate flux function that calculates $\hat{\nF}_{\xi^l}$
and many approaches can be found in the literature. However, the specific
choice is crucial for the accuracy of the numerical solution. This is further
discussed in \cref{sec:numfluxes} in the context of flows at low Mach numbers
\begin{equation}
	M=\frac{|\vec v|}{c},
\end{equation}
where $c$ is the \emph{speed of sound} given by \cref{def:speedOfSound}.

The values that enter the approximate flux function need to be reconstructed
from the center of the cells to the corresponding interfaces. The
reconstruction and the evaluation of the flux is done for each coordinate
direction separately, before the resulting fluxes over the surfaces are added
for each cell.

The semi-discrete scheme is then evolved in time using an ODE solver, such as
a Runge--Kutta method. With an at least linear reconstruction and
a sufficiently accurate ODE solver this discretization yields a second-order
accurate scheme as has been numerically verified by \cite{berberich2019a}.
For the tests in this article we mainly use the implicit second-order accurate
three step Runge--Kutta method ESDIRK23 of \citet{hosea1996a}.

We chose an advective CFL time step
($\text{CFL}_u$), not strictly for reasons of stability, but as a good
compromise between accuracy and efficiency. In curvilinear coordinates it takes
the form
\begin{equation}
  \Delta t_{\text{CFL}_u} = c_\text{CFL} \min_{ijkl} \frac{(\Delta \xi^l)_{ijk}}{|\vec n_{\xi^l} \cdot \vel|_{ijk}},
\end{equation}
with a constant $c_\text{CFL}$ of order unity and an estimate of the cell
length in direction~$\xi^l$ given by
\begin{equation}
  (\Delta \xi^1)_{ijk} = \frac{V_{ijk}}{\half \left(A_{i-\half,j,k}+A_{i+\half,j,k}\right)},
\end{equation}
and accordingly for $(\Delta \xi^2)_{ijk}$ and $(\Delta \xi^3)_{ijk}$.

This time step criterion generally works well when the flow is fully developed,
but it has problems when the Mach numbers on the grid are very small (e.g., in
the beginning of a simulation with zero initial velocities), because this
yields very large or infinite time steps. As a way to prevent this,
\citet{miczek2013a} suggests to include the free-fall signal velocity in the
time step calculation. The so-called $\text{CFL}_{ug}$ time step is then given
by
\begin{equation}
  \Delta t_{\text{CFL}_{ug}} = c_\text{CFL} \min_{ijkl}
  \frac{(\Delta \xi^l)_{ijk}}{s_{ijkl}},
  \label{eq:cflug}
\end{equation}
with the signal velocity
\begin{equation}
  s_{ijkl} = \half \left( a^l |\vec n_{\xi^l} \cdot \vel| +
  \sqrt{|\vec n_{\xi^l} \cdot \vel|^2 + 4 a^l c_\text{CFL} (\Delta \xi^l) \vec n_{\xi^l} \cdot \vec g} \right)_{ijk}.
  \label{eq:cflug-s}
\end{equation}
The parameter~$a^l$ selects the right branch of the quadratic solution and is given in \cref{tab:cflug-a}.
\begin{table}
  \caption{\label{tab:cflug-a}Parameter~$a^l$ used in the computation of the signal velocity in \cref{eq:cflug-s}.}
  \begin{tabular}{l|c}
    condition & $a^l$\\
    \hline
    $\vec n_{\xi^l} \cdot \vel > 0$, $\vec n_{\xi^l} \cdot \vec g > 0$ & $+1$ \\
    $\vec n_{\xi^l} \cdot \vel \leq 0$, $\vec n_{\xi^l} \cdot \vec g \leq 0$ & $-1$ \\
    $\vec n_{\xi^l} \cdot \vel > 0$, $\vec n_{\xi^l} \cdot \vec g \leq 0$,
    $\frac{\left(\vec n_{\xi^l} \cdot \vel\right)^2}{4 \vec n_{\xi^l} \cdot \vec g} + c_\text{CFL} (\Delta \xi^l) \leq 0$ & $+1$ \\
    $\vec n_{\xi^l} \cdot \vel > 0$, $\vec n_{\xi^l} \cdot \vec g \leq 0$,
    $\frac{\left(\vec n_{\xi^l} \cdot \vel\right)^2}{4 \vec n_{\xi^l} \cdot \vec g} + c_\text{CFL} (\Delta \xi^l) > 0$ & $-1$ \\
    $\vec n_{\xi^l} \cdot \vel \leq 0$, $\vec n_{\xi^l} \cdot \vec g > 0$,
    $\frac{\left(\vec n_{\xi^l} \cdot \vel\right)^2}{4 \vec n_{\xi^l} \cdot \vec g} + c_\text{CFL} (\Delta \xi^l) > 0$ & $-1$ \\
    $\vec n_{\xi^l} \cdot \vel \leq 0$, $\vec n_{\xi^l} \cdot \vec g > 0$,
    $\frac{\left(\vec n_{\xi^l} \cdot \vel\right)^2}{4 \vec n_{\xi^l} \cdot \vec g} + c_\text{CFL} (\Delta \xi^l) \leq 0$ & $+1$ \\
  \end{tabular}
\end{table}

For Mach numbers close to $M=1$, however, it is usually more efficient to use
explicit time stepping. For this we use the third-order accurate RK3 scheme of
\cite{shu1988a} with a $\text{CFL}_{uc}$ time step controlled by the fluid
velocity and sound speed. It is given by
\begin{equation}
  \Delta t_{\text{CFL}_{uc}} = \frac{c_\text{CFL}}{N_\text{dim}} \min_{ijkl} \frac{(\Delta
\xi^l)_{ijk}}{|\vec n_{\xi^l} \cdot \vel|_{ijk} + c_{ijk}},
\label{eq:cfluc}
\end{equation}
where $N_\text{dim}$ denotes the spatial dimensionality of the equations. In
contrast to the previous criteria, this is a strict stability criterion for the
explicit time stepping. We note that the use of a third-order scheme is not
strictly necessary for the presented results. A second-order time integration
scheme, such as RK2 \citep{shu1988a}, yields virtually identical results in
combination with second-order spatial reconstruction.

\subsection{Numerical flux functions}
\label{sec:numfluxes}
\label{sec:ausm}
A fundamental part of the discretization is
the choice of a numerical two-state flux. These fluxes give approximate
solutions of the two-state Riemann problem at the cell interfaces. Choosing
different numerical fluxes yields different properties for the scheme.
Many of the typically used Riemann solvers or other flux functions suffer from
excessive Mach number dependent diffusion. In the case of the Roe solver
\citep{roe1981a} the origin of this is an upwind term in the schemes that is
needed for numerical stability
\citep[e.g.,][]{turkel1987a,guillard1999a,miczek2015a}. Other Godunov-type
schemes are subject to similar issues \citep{guillard2004a}. To correct this
behavior, a number of low Mach number fixes have been proposed that aim on
reducing the excessive diffusion and make it independent of the Mach number
\citep[e.g.,][]{turkel1987a,li2008a,rieper2011a,osswald2015a,miczek2015a,barsukow2017a,berberich2020b}.

One peculiarity of astrophysical setups compared to, for example, setups in the
engineering community is the presence of strong stratifications where pressure
and density may change by orders of magnitudes within the computational domain.
In such setups, the reduction of diffusion comes with the risk of
reducing stability and many of the schemes found in the literature develop
instabilities. The \slh code is designed in a modular fashion that facilitates
the implementation and testing of different types of flux functions. In
numerical tests we find the so-called \ausmpup method to yield appropriate
results in the low-Mach regime in combination with the well-balancing method
discussed here. The basic construction and a modification for improved low-Mach
behavior is discussed here.

An approach to numerical flux functions that can easily be extended to flows at
low Mach numbers is followed in the class of \emph{Advective Upstream Splitting
Methods} (AUSM), which have been first introduced by \citet{liou1993a}. In
\citet{liou1996a} the AUSM scheme was extended to AUSM${}^+$, the idea of which
we briefly describe in the following. To be consistent with the original
publication, we use dimensionalized quantities.

The central idea is to split the analytical flux function $\vec f_\chi$ of
\cref{def:euler3d} into a pressure and a mass flux via
\begin{equation}
\vec f_{x_i} = p\,\vec e_{i+1} + \dot m_i \psi,
\label{eq:AUSM}
\end{equation}
with
\begin{equation}
\dot m_i = \rho v_i,\,\quad
\psi =
\begin{pmatrix}
1 \\ u\\v\\w\\E+\frac{p}{\rho}
\end{pmatrix},\quad i\in[1,2,3]
\end{equation}
and the $i$-th canonical basis vector in the five-dimensional flux vector space
$\vec e_i$. This formulation is given for Cartesian
coordinates, a transformation to curvilinear coordinates is possible.

The pressure and mass flux of \cref{eq:AUSM} are discretized separately which
results in the numerical flux function
\begin{align}
   \hvF_{x_i}(\vec U_L,\vec U_R) = p_{1/2}(\vec U_L,\vec U_R)\,\vec e_{i+1} + \dot m_{1/2}(\vec U_L,\vec U_R)\psi_{\text{up}}(\vec U_L,\vec U_R),
\end{align}
where the upwind term $\psi_{\text{up}}$ is given by
\begin{equation}
\psi_\text{up}(\vec U_L,\vec U_R)=
\left\{
\begin{array}{ll}
\psi(\vec U_L) &\text{ if }\dot m_{1/2}(\vec U_L,\vec U_R) \geq 0,\\
\psi(\vec U_R) &\text{ otherwise}.
\end{array}
\right.
\end{equation}
The core properties of this numerical flux function are determined by the
definition of the interface values $p_{1/2}$ and $\dot m_{1/2}$ of the pressure
$p$ and the mass flux $\dot m_i$. With the initially proposed definitions of
\cite{liou1996a}, this flux function is not capable of resolving low Mach
number flows. However, \citet{liou2006a} extended the AUSM$^+$ scheme to
\ausmpup with enhanced low Mach number capability.

For \ausmpup, the interface pressure is defined as
\begin{align}
  p_{1/2} &= \mathcal{P}^+_{(5)}(M_L)p_L + \mathcal{P}^-_{(5)}(M_R)p_R \nonumber\\
          &\phantom{=}- K_u \mathcal{P}^+_{(5)}(M_L) \mathcal{P}^+_{(5)}(M_R) (\rho_L + \rho_R) (f_a\, c_{1/2}) (u_R - u_L),
  \label{eq:p12}
\end{align}
where $\mathcal{P}^\pm_{(5)}$ are fifth degree polynomial functions, $c_{1/2}$
is an approximation for the interface speed of sound, and $K_u$ is a constant that
can be set to a value between zero and unity. We refer the reader to
\citet{liou2006a} for the detailed definitions of the terms. The third term on
the right hand side of \cref{eq:p12} that includes velocity-diffusion is called
$u$-term and is designed to reduce the numerical dissipation at low Mach
numbers. It involves a scaling factor $f_a$ defined as
\begin{equation}
f_a=\Mo (2-\Mo),
\end{equation}
with
\begin{equation}
\Mo=\min\left[1,\max\left(\ma,\Mcut \right)\right],
\label{eq:mo}
\end{equation}
where $\Mcut$ is a cut-off Mach number that ensures that $f_a$ does not
approach 0 in the limit of very small Mach numbers. This is necessary to
prevent singularities as the inverse of $f_a$ enters into the mass-flux part
[see \cref{eq:ausmpupmass}] in the original definition of \ausmpup. However, as
described below, the \slh code sets the scaling in these two parts
independently such that $\Mcut$ can be theoretically set to zero. In \slh, for
implementation reasons the value is set to a small value, typically to
\num{e-13}, to avoid divergence at smaller Mach numbers. This could easily be
changed, but does not have a practical influence on our calculations. The mass
flux in \ausmpup is given by
\begin{equation}
  \dot{m}_{1/2} = c_{1/2} M_{1/2}
  \left\{
    \begin{array}{ll}
      \rho_L &\text{ if }\dot M_{1/2} > 0,\\
      \rho_R &\text{ otherwise},
    \end{array}
  \right.
  \label{eq:ausmpupmass}
\end{equation}
with the interface Mach number
\begin{equation}
  M_{1/2} = \mathcal{M}^+_{(4)}(M_L) + \mathcal{M}^-_{(4)}(M_R) - \frac{K_p}{\fapdiff} \max \left(1 - \sigma \bar{M}^2,0\right) \frac{p_R - p_L}{\rho_{1/2} c^2_{1/2}}.
\end{equation}
Here, $\mathcal{M}^\pm_{(4)}$ are fourth degree polynomial functions and
$\bar{M}=(u_L^2 + u_R^2)/(2c_{1/2}^2)$. $K_p$ and $\sigma$ are constants
between zero and unity. The last term is called the $p$-term and is a pressure
diffusion term that was introduced to ensure pressure--velocity coupling at low
speeds \citep{edwards1998a}. In the original \ausmpup scheme the Mach number
dependent scaling of the $p$-term, $\fapdiff$, was chosen identical to that of
the $u$-term, $f_a$. Similar to \citet{miczek2013a} we chose independent cut-off
values for $f_a$ and
\begin{align}
\fapdiff =M_o^p (2-M_{o}^p),\quad M_o^p=\min\left[1,\max\left(M,\Mpcut\right)\right].
\end{align}
This way, $\fapdiff$ can be defined with a significantly higher cut-off Mach
number (typically around \num{e-1}) compared to $f_a$. This prevents stability
issues, which can occur at locally very low Mach numbers when $1/\fapdiff$
becomes exceedingly large. We use this modified scheme in the presented tests
and refer to it as \ausm. Still even this scheme is struggling with maintaining
a simple hydrostatic stratification as shown in \cref{sec:hystattest}.

It is important to note that the role of the effect of the pressure diffusion
is altered by combining \ausm with any of the \wbing methods described in the
following: Well-balancing techniques lead to an exact reconstruction of the
hydrostatic pressure. Only the nonhydrostatic pressure deviations are captured
by the reconstruction and given to the numerical flux. Hence, when \wbing is
applied, the pressure diffusion acts on the nonhydrostatic pressure only.

We define a corresponding basic scheme called \ausmp by setting $\Mcut=\Mpcut=1$. This scheme does have high dissipation at low Mach numbers and we just use it to assess the interaction of the various \wbd schemes with low Mach number flux functions.

\section{Well-balancing methods}
To illustrate the issue with configurations close to hydrostatic equilibrium in finite-volume codes we consider the effect of one time step on an initial configuration in perfect hydrostatic equilibrium. For simplicity we discretize the time derivative using the forward Euler method and only consider the one-dimensional Euler equations. From the analytic result we expect that this step will not alter the states and any subsequent steps will also keep the hydrostatic equilibrium intact. To this end we associate a discrete stationary solution that provides a good approximation to the hydrostatic equilibrium. If starting from a discrete hydrostatic equilibrium, the solution of the time evolutionary problem does not change for a numerical scheme, we call it \emph{\wbd}.

The density, momentum, and total energy after one time step of length $\Delta t$ (denoted by the superscript ``1'') are calculated from the previous values (denoted by the superscript ``0''), the interface fluxes $\hat{\vec F}$, and the cell-centered source term $\hat{\vec S}$. The result for cell~$i$ is
\begin{align}
  \rho_i^1 &= \rho_i^0 - \frac{\Delta t}{\Delta x}\left[ \left(\hat{\vec F}_{i+\half}^0\right)_1 - \left(\hat{\vec F}_{i-\half}^0\right)_1\right],\\
  (\rho u)_i^1 &= (\rho u)_i^0 - \frac{\Delta t}{\Delta x}\left[ \left(\hat{\vec F}_{i+\half}^0\right)_2 - \left(\hat{\vec F}_{i-\half}^0\right)_2\right] + \Delta t \left(\hat{\vec S}_i\right)_2,\\
  E_i^1 &= E_i^0 - \frac{\Delta t}{\Delta x}\left[ \left(\hat{\vec F}_{i+\half}^0\right)_3 - \left(\hat{\vec F}_{i-\half}^0\right)_3\right],
\end{align}
where the indexes outside the parentheses stand for the vector component of the flux or source term. To be \wbd, a scheme needs to guarantee that the step leaves the state unchanged, which leads to the conditions
\begin{align}
  \label{eq:wb-1}0 &= \left(\hat{\vec F}_{i+\half}^0\right)_1 - \left(\hat{\vec F}_{i-\half}^0\right)_1,\\
  \label{eq:wb-2}0 &= \left(\hat{\vec F}_{i+\half}^0\right)_2 - \left(\hat{\vec F}_{i-\half}^0\right)_2 - \Delta x \left(\hat{\vec S}_i\right)_2,\\
  \label{eq:wb-3}0 &= \left(\hat{\vec F}_{i+\half}^0\right)_3 - \left(\hat{\vec F}_{i-\half}^0\right)_3.
\end{align}
As the fluxes are evaluated at different states, none of these conditions is fulfilled automatically. A consistent numerical flux will automatically satisfy \cref{eq:wb-1,eq:wb-3} because the 1- and 3-components of the fluxes are zero for $u=0$. The case of \cref{eq:wb-2} is less straightforward. Here, the discretization of the source term must be constructed to match the flux difference in the hydrostatic case. The methods described in the following subsections achieve this in different ways.

\subsection{\leroux method}
\label{sec:clr}
\subsubsection{The one-dimensional \leroux method}
\label{sec:cargo-1d}
One method to turn almost any hydrodynamics scheme into a well-balanced
scheme was suggested by \citet{cargo1994a} \citep[see also][]{leroux1999a}.
The only prerequisites it needs are support for a general equation of state
and flux functions that resolve contact discontinuities. For completeness
we describe the original method, which is only applicable in the
one-dimensional case, and turn to the multidimensional extension in
\cref{sec:multi-d}. The one-dimensional Euler equations with gravity read
\begin{equation}
  \label{eq:euler1d}
  \DD{}{t} \begin{pmatrix}
    \rho\\
    \rho u\\
    E'
  \end{pmatrix} +
  \DD{}{x} \begin{pmatrix}
    \rho u\\
    \rho u^2 + p\\
    u (E' + p)\\
  \end{pmatrix}
  = \begin{pmatrix}
    0\\
    \rho g\\
    \rho g u\\
  \end{pmatrix},
\end{equation}
with $g$ being the, possibly negative, gravitational acceleration in the
$x$-direction. The 1D method presented here relies on gravity being constant
in space and time. Furthermore, $E'=\rho\epsilon+\half \rho|\vel|^2$ is the
total energy excluding potential energy. It is only used in the one-dimensional
method in this subsection. The multidimensional extension in \cref{sec:multi-d}
follows a slightly different principle using the total energy $E$ including the
potential as defined in \cref{sec:equations}.

\citet{cargo1994a} suggest the introduction of a potential~$q$ defined
by its spatial and temporal derivatives
\begin{equation}
  \DD{q}{x} = \rho g, \quad \DD{q}{t} = - \rho u g.
    \label{eq:q-def}
\end{equation}

Numerically, this potential is treated like a composition variable,
meaning that its time evolution is determined by the advection equation,
\begin{equation}
  \DD{(\rho q)}{t} + \DD{(\rho q u)}{x} = 0.
\end{equation}
This ensures that the conditions of Eq.~\eqref{eq:q-def} are fulfilled at
all times if they are satisfied initially.

Expressing the right side of Eq.~\eqref{eq:euler1d} using $q$
yields
\begin{align}
  \label{eq:cargo-euler-2}
  \DD{}{t} \begin{pmatrix}
    \rho\\
    \rho u\\
    E'
  \end{pmatrix} +
  \DD{}{x} \begin{pmatrix}
    \rho u\\
    \rho u^2 + p\\
    u (E' + p)\\
  \end{pmatrix}
  &= \begin{pmatrix}
    0\\
    \DD{q}{x}\\
    - \DD{q}{t}\\
  \end{pmatrix}.\\
  \intertext{Collecting derivatives with respect to the same variable and inserting $0 = q - q$ results in}
  \label{eq:cargo-euler-3}
  \DD{}{t} \begin{pmatrix}
    \rho\\
    \rho u\\
    E' + q
  \end{pmatrix} +
  \DD{}{x} \begin{pmatrix}
    \rho u\\
    \rho u^2 + p - q\\
    u (E' + q + p - q)\\
  \end{pmatrix}
  &= \begin{pmatrix}
    0\\
    0\\
    0\\
  \end{pmatrix}.
\end{align}
At this point we introduce a modified \eos based on an arbitrary original \eos
by defining a new pressure, $\Pi$, and total energy per volume, $F$, as
\begin{equation}
  \Pi = p - q, \quad F' = E' + q.
  \label{eq:cargo-eos}
\end{equation}
With this definition \cref{eq:cargo-euler-3} takes the form of the homogeneous
Euler equations for a modified \eos,
\begin{align}
  \label{eq:cargo-euler-4}
  \DD{}{t} \begin{pmatrix}
    \rho\\
    \rho u\\
    F'
  \end{pmatrix} +
  \DD{}{x} \begin{pmatrix}
    \rho u\\
    \rho u^2 + \Pi\\
    u (F' + \Pi)\\
  \end{pmatrix}
  &= \begin{pmatrix}
    0\\
    0\\
    0\\
  \end{pmatrix}.
\end{align}
The physical meaning of $q$ is that of hydrostatic pressure, except for an
arbitrary constant offset. This means that the modified pressure~$\Pi$ of
an atmosphere in perfect hydrostatic equilibrium is spatially constant,
making the solution of the Euler equations trivial.

This new \eos does not change the speed of sound. This can easily be seen from
rewriting the expression for the speed of sound for a general \eos
\eqref{def:speedOfSound} in terms of $\Pi$ and $F'$. Because $q$ does not
depend on any other thermodynamic variables, the rewritten expression takes the
same form as the original.

\subsubsection{A multidimensional extension}
\label{sec:multi-d}
An obvious multidimensional extension of the aforementioned method would
be introduce a potential~$q$ with the defining properties
\begin{equation}
  \label{eq:q-multi-d-1}
  \nabla q = \rho \vec{g}, \quad \DD{}{t} q = - \rho \vec{g} \cdot
  \vec{u}.
\end{equation}
While the algebra shown in
\cref{eq:cargo-euler-2,eq:cargo-eos,eq:cargo-euler-3,eq:cargo-euler-4}
is still valid for this new potential, the problem with this definition is
the mere existence of this potential~$q$. From \cref{eq:q-multi-d-1} follows that
\begin{equation}
  \nabla\times\left(\nabla q\right) = \nabla \times \rho \g = \vec{0},
\end{equation}
and using the fact that $\nabla \times \vec{g}=0$, due to $\vec{g}$ being
derived from a gravitational potential, we can simplify this equation to
\begin{equation}
  \nabla \times \rho \vec{g} = \rho \nabla \times \vec{g} - \vec{g}
  \times \nabla \rho = - \vec{g} \times \nabla \rho=\vec{0}.
  \label{eq:curlzero}
\end{equation}
This means that a potential only exists if the cross product $\vec{g} \times
\nabla \rho$ vanishes. This only happens for any of the following three
conditions: the trivial case of gravity being globally zero, the case of
constant density, and the case where the gradient of density has the same
equilibrium with its surroundings. Thus, the approach of \cref{eq:q-multi-d-1}
is not suitable.

In order to construct a general, but approximate, multidimensional extension of
the well-balanced method from \cref{sec:cargo-1d}, we restrict ourselves to
problems in which $\rho$ does not vary strongly on equipotential surfaces of
the gravitational potential $\phi$.  We denote an average on these
equipotential surfaces, which we call \emph{horizontal average}, with the
operator $\avg{\cdot}$, so we can define an averaged density
\begin{equation}
  \label{eq:def-rho0}
  \rho_0 = \avg{\rho}.
\end{equation}
By definition of the horizontal average, the gradient of $\rho_0$ is
parallel to $\vec{g}$. This allows us to define the potential by
\begin{equation}
  \label{eq:def-q}
  \nabla q = \rho_0 \vec{g}, \quad \DD{}{t} q = 0,
\end{equation}
for which \cref{eq:curlzero} is fulfilled automatically.

The fluxes (\cref{def:fluxVectorTransformed}) and the source term
(\cref{def:source}) in the compressible Euler equations (\cref{def:euler3d})
can then be rewritten as
\begin{equation}
\f_{\xi^l } =
\begin{pmatrix}
\rho \, \vec n_{\xi^l}^T \vel \\
\rho u \,\vec n_{\xi^l}^T \vel + (n_{\xi^l})_{x}\Pi\\
\rho v \,\vec n_{\xi^l}^T \vel + (n_{\xi^l})_{y}\Pi\\
\rho w \,\vec n_{\xi^l}^T \vel + (n_{\xi^l})_{z}\Pi\\
\vec n_{\xi^i}^T \vel \left( F + \Pi \right)
\end{pmatrix},
\quad
\vec \s =
\begin{pmatrix}
  0\\-(\rho-\rho_0)\DD{\phi}{x}\\-(\rho-\rho_0)\DD{\phi}{y}\\-(\rho-\rho_0)\DD{\phi}{z}\\
  0
\end{pmatrix}.
\end{equation}
There are two
fundamental differences between this form and \cref{eq:cargo-euler-4}. First,
$F=E + q$ is defined including the potential energy. This is different
from the one-dimensional case in \cref{eq:cargo-eos}. $q$ is now temporally
constant, but because of the different definition of the total energy, the
source term in the energy equation still vanishes. Second, the source term in
the momentum equation does not completely vanish anymore. It is now
proportional to the local deviation of density from its horizontal average.
Even though this scheme loses some of the advantageous properties of the
one-dimensional version, it is still a significant improvement for
multidimensional simulations of stratified atmospheres.

A positive side effect of $q$ being temporally constant is that, in contrast to
the original \leroux method, the gravitational acceleration, $g$, is now
allowed to vary spatially. A temporal variation of $g$ is also possible, for
example through self-gravity, but that necessitates a recomputation of $q$ to
keep the \wbd property.

Because it only requires to a slight modification of the \eos, it is expected
that the \leroux \wbing method can be implemented into an existing hydrodynamic
code rather easily, provided it supports a general \eos. \leroux \wbing
generally only works with flux functions that preserve contact discontinuities.
It works well with the \ausmpup family of fluxes used in this paper.

\subsection{The \alphabeta method}
\label{sec:alphabeta}

Another approach to balance any hydrostatic solution is the \alphabeta \wbing
method presented by \cite{berberich2018a}. Similar to the \leroux method, the
hydrostatic solution needs to be known a priori. An advantage of the \alphabeta
\wbing however is, that it permits an arbitrary, even multidimensional,
structure of the hydrostatic target solution.  Hence, the \alphabeta method is
able to deal with the example of a low-density bubble in pressure-equilibrium
with its surrounding of \cref{sec:clr} that could not be balanced with the
\leroux method.

The key idea of this method is to replace the physical values of $\rho$ and $p$
by their respective relative deviation prior to the reconstruction at the cell
interfaces. The reconstructed values are then multiplied by the known
hydrostatic solution at the interfaces. The second-order approximation of gravity
is constructed such that the source term exactly cancels the flux over the interface and
hence that the initial hydrostatic stratification is maintained to machine
precision.

For convenience, the working principle of the \alphabeta\wbing method is
demonstrated for the one-dimensional Euler equations in Cartesian coordinates.
The extension to higher dimensional curvilinear grids follows the same
principle and can be found in \cite{berberich2019a} together with a
mathematically more rigorous formulation of the method.

For a given gravitational potential $\phi(x)$, we denote $\tilde\rho$ and
$\tilde p$ as the solution to the hydrostatic equation [\cref{eq:hystat}] in
one dimension, that is
\begin{equation}
  \frac{\partial \tilde p}{\partial x}=-\tilde\rho \df{\phi(x)}{x}.
  \label{eq:hystatabeta}
\end{equation}
The solutions are written as
\begin{equation}
  \tilde{\rho}=\rho_0\,\alpha(x), \quad \tilde p=p_0\,\beta(x),
  \label{eq:defabeta}
\end{equation}
where $\alpha(x),\,\beta(x)$ are dimensionless profiles and $\rho_0,\,p_0$ carry the
physical dimension of density and pressure, respectively. It is assumed that
the profiles in \cref{eq:defabeta} are known at least at coordinates that
coincide with cell centers and interfaces.

The numerical solution of the one-dimensional Euler equation in the finite
volume approach requires the reconstruction of the quantities from each cell
center $i$ of the computational domain to the respective cell interfaces
$i+\frac12$. In general, there is some freedom in choosing the set of
quantities that is reconstructed in addition to the velocities. This is
exploited by the \alphabeta\wbing method, which considers the relative
deviation from the hydrostatic solution. The set of quantities at cell $i$ that
are reconstructed is hence chosen to be
\begin{equation}
  \vec{W}_i =  \left(W_i^{\rho},W_i^u,W_i^p\right) =
  \left(
    \frac{\rho_i}{\alpha(x_i)},
    u_i,
    \frac{p_i   }{\beta (x_i)}
  \right).
  \label{eq:rec_vars-1d}
\end{equation}
There is no restriction on the specific choice of the reconstruction scheme
that calculates the value of $W_i$ at the interface, that is $W_{i\pm\frac12}$.
After reconstruction, the variables are transformed back to their physical
counterpart by multiplication with the known hydrostatic solution at the
interfaces,
\begin{align}
  \rho_{i+\frac12}^{L/R} &= \alpha\left(x_{i+\frac12}\right)\left(W_{i+\frac12}^\rho\right)^{L/R},&
  u_{i+\frac12}^{L/R} &= \left(W_{i+\frac12}^u\right)^{L/R},\nonumber\\
  p_{i+\frac12}^{L/R} &=\beta\left(x_{i+\frac12}\right) \left(W_{i+\frac12}^p\right)^{L/R}.
  \label{eq:abrec}
\end{align}
Here, $L/R$ denotes the value at the interface when reconstructing from the
left or right side, respectively. The values given by \cref{eq:abrec} enter
into the numerical flux function (see \cref{sec:numfluxes}).

If density $\rho$ and pressure $p$ on the computational grid correspond to
the hydrostatic solution \cref{eq:defabeta} and $u\equiv 0$, it follows that
the quantities reconstructed from the left and right side are equal for all
cell interfaces and hence
\begin{align}
\vec U_{i+\frac12}^L=\vec U_{i+\frac12}^R=\vec U_{i+\frac12}.
\end{align}
If the left and right interface state are the same, any numerical flux function
has to equal the analytical flux function to ensure the consistency of the
method. Thus,
\begin{align}
  \hat{\vec{F}}_x\left(\vec U_{i+\frac12}^L,\vec U_{i+\frac12}^R\right)
  = \vec f_x\left(\vec U_{i+\frac12}\right)
  = \begin{pmatrix}
     0 \\ \tilde p(x_{i+\frac12}) \\ 0
    \end{pmatrix},
    \label{eq:reflux}
\end{align}
which immediately follows from \cref{def:fluxVectorTransformed} for vanishing
velocities.

In order to maintain hydrostatic equilibrium, the residual flux of
\cref{eq:reflux} has to be balanced exactly by the source term $\s$ in
\cref{def:euler3d}. To achieve this, the \alphabeta method expresses the
gravitational potential in \cref{def:source} with the aid of the hydrostatic
equation \cref{eq:hystatabeta} as
\begin{align}
    -\df{\phi}{x} = \frac{p_0}{\rho_0\,\alpha(x)}\df{\beta(x)}{x}.
\end{align}
The one-dimensional source term for gravity [see \cref{eq:sourceterm_1d}] is
then given by
\begin{align}
    \hat{\vec S}_{i}=\begin{pmatrix}0\\s_i\\0\end{pmatrix},\quad
    s_i = \frac{p_0}{\rho_0} \frac{\beta_{i+\frac12} - \beta_{i-\frac12}}{\Delta x} \frac{\rho_i}{\alpha_i},
\label{eq:abetasource}
\end{align}
which is a second-order accurate discretization.  If the states on the
computational grid correspond to the hydrostatic solution, then
\cref{eq:abetasource} reduces to
\begin{align}
	\s_i &= \begin{pmatrix}0\\\tilde p\left(x_{i+\frac12}\right)-\tilde p\left(x_{i-\frac12}\right)\\0\end{pmatrix} \nonumber\\
	&= \hat{\vec F}_x\left(\vec U_{i+\frac12}^L,\vec U_{i+\frac12}^R\right) - \hat{\vec F}_x\left(\vec U_{i-\frac12}^L,\vec U_{i-\frac12}^R\right)
	\label{eq:wbproperty}
\end{align}
and the discretized source term exactly cancels the interface fluxes. This
leads to zero residual and thus to a well-balanced scheme. The well-balanced
property for the one-dimensional \alphabeta method is formally shown in
\citet{berberich2018a}.

\subsection{The deviation method}
\label{sec:deviation}
In the following we give a short description of the simple and general well-balanced method introduced in \cite{berberich2021a}. For more details we refer the reader to this reference.
The core of the method is the \emph{target solution} $\tilde\u$, which must be known a priori. It has to be a stationary solution to the Euler equations \eqref{def:euler3d}, that is it has to satisfy the relation
\begin{equation}
	\label{eq:euler3d_stationary}
	A_{\xi}\frac{\del\f_{\xi}(\tilde{\u})}{\del \xi} + A_{\eta}\frac{\del\f_{\eta}(\tilde{\u})}{\del \eta} + A_{\zeta}\frac{\del\f_{\zeta}(\tilde{\u})}{\del \zeta} = J \s(\tilde{\u}).
\end{equation}
It is noteworthy that, in contrast to other \wbing methods, it can include a nonzero velocity. In the numerical applications in \cref{sec:tests} we are going to store the hydrostatic or stationary solution which shall be well-balanced in $\tilde{\u}$.
Subtracting \cref{eq:euler3d_stationary} from \cref{def:euler3d} yields the evolution equation
\begin{align}
\label{eq:euler3d_for_deviations}
	J\frac{\del (\Delta\u)}{\del t}
	+ & A_{\xi}\left( \frac{\del\f_{\xi}(\tilde \u+\Delta\u)}{\del \xi}-\frac{\del\f_{\xi}(\tilde\u)}{\del \xi} \right) \nonumber\\
	+ & A_{\eta}\left(\frac{\del\f_{\eta}(\tilde \u+\Delta\u)}{\del \eta}-\frac{\del\f_{\eta}(\tilde\u)}{\del \eta}\right)\nonumber \\
	+ & A_{\zeta}\left(\frac{\del\f_{\zeta}(\tilde \u+\Delta\u)}{\del \zeta}-\frac{\del\f_{\zeta}(\tilde\u)}{\del \zeta}\right) = J \left(\s(\tilde \u+\Delta \u)- \s(\tilde \u)\right)
\end{align}
for the deviations $\Delta \u=\u-\tilde{\u}$ from the target solution $\tilde{\u}$. In order to obtain a well-balanced scheme which exactly maintains the stationary target solution $\tilde{\u}$, we discretize \cref{eq:euler3d_for_deviations} instead of \cref{def:euler3d}. This yields
\begin{align}
\label{eq:finiteVolumeScheme_deviation}
V_{ijk}\D[(\Delta\vec U)_{ijk}]{t} = -\,
&     A_{i+\half,j,k}(\hat{\vec F}^{\rm dev}_\xi  )_{i+\half,j,k}
+ A_{i-\half,j,k}(\hat{\vec F}^{\rm dev}_\xi  )_{i-\half,j,k}\nonumber\\
-\, & A_{i,j+\half,k}(\hat{\vec F}^{\rm dev}_\eta )_{i,j+\half,k}
+ A_{i,j-\half,k}(\hat{\vec F}^{\rm dev}_\eta )_{i,j-\half,k} \nonumber\\
-\, & A_{i,j,k+\half}(\hat{\vec F}^{\rm dev}_\zeta)_{i,j,k+\half}
+ A_{i,j,k-\half}(\hat{\vec F}^{\rm dev}_\zeta)_{i,j,k-\half} \nonumber\\
+\,&V_{ijk}\hat{\vec S}^{\rm dev}_{ijk},
\end{align}
where the numerical flux differences
\begin{align*}
	\left(\hat{\nF}^{\rm dev}_\xi\right)_{i+\half,j,k} =& \left(\hat{\nF}_\xi\right)_{i+\half,j,k} - \f_\xi\left[\tilde{\u}\left(\x_{i+\half,j,k}\right)\right],\\
	\left(\hat{\nF}^{\rm dev}_\eta\right)_{i,j+\half,k} =& \left(\hat{\nF}_\eta\right)_{i,j+\half,k} - \f_\eta\left[\tilde{\u}\left(\x_{i,j+\half,k}\right)\right],\\
	\left(\hat{\nF}^{\rm dev}_\zeta\right)_{i,j,k+\half} =& \left(\hat{\nF}_\zeta\right)_{i,j,k+\half} - \f_\zeta\left[\tilde{\u}\left(\x_{i,j,k+\half}\right)\right]
\end{align*}
are the differences between the numerical fluxes evaluated at the states
$\tilde{\u}+\Delta \vec U^{L/R}$ and the exact fluxes evaluated at the values
of the target solution at the interface centers. The interface values $\Delta
\vec U^{L/R}$ are obtained via reconstruction. To reconstruct in the set of
variables $\u^\text{other}=\mathcal T \left(\u \right)$ that is different from
conserved variables $\u$, the reconstruction is applied to the transformed
deviations
\begin{equation}
\Delta \nU_{ijk}^\text{other}=\mathcal T\left(\tilde{\u}\left(\x_{ijk}\right)+\Delta \nU_{ijk}\right)-\mathcal T\left(\tilde{\u}\left(\x_{ijk}\right)\right).
\end{equation}
After reconstruction, the interface values are transformed back. For the left
interface values, this reads
\begin{equation}
	\nU_{i+\half,j,k}^{L} =\mathcal T^{-1} \left(\mathcal T\left(\tilde\u\left(\x_{i+\half,j,k}\right)\right)+\Delta\nU_{i+\half,j,k}^\text{other}\right),
\end{equation}
and right interface values are calculated likewise.  The source term difference
discretization in \cref{eq:finiteVolumeScheme_deviation} is
\begin{equation}
\hat{\vec S}^{\rm dev}_{ijk} = \hat{\vec S}\left(\tilde \u\left(\x_{ijk}\right)+\Delta\nU_{ijk}\right)
- \hat{\vec S}\left(\tilde \u\left(\x_{ijk}\right)\right),
\end{equation}
where $\hat{\nS}(\nU)$ means that the source term discretization \cref{eq:sourceterm_1d} is evaluated at the state $\nU$.
It has been shown in \cite{berberich2021a} that this modification of the scheme
\eqref{eq:finiteVolumeScheme} renders it well-balanced in the sense that the residual vanishes if $\u=\tilde{\u}$ and that the method can be applied in arbitrarily high order finite-volume codes.

This method follows ideas already published elsewhere in the literature. \citet{veiga2019a} introduce a similar \wbd scheme in the context of finite element methods. The method introduced in \citet{dedner2001a} for stratified MHD flows with gravity also shares many features with the deviation method. The key difference is that the \citet{dedner2001a} method subtracts the residual of the initial state of the simulation, while the deviation method subtracts a known background state during the spatial reconstruction step, which is at an earlier stage. While we have not quantified this in tests, it is expected that the deviation method will produce a more accurate reconstruction close to equilibrium because it is essentially reconstructing a constant function.

\section{Numerical tests and application examples}
\label{sec:tests}

This section presents simple test simulations that verify that the presented
\wbing methods in combination with a low-Mach flux function are stable and do
enable correct representation of slow flows in stellar-type stratifications.
This is done for steady-state and dynamical test problems.  Unless stated
otherwise, time integration is performed with the implicit ESDIRK23 scheme in
combination with the $\text{CFL}_{ug}$ time step criterion [\cref{eq:cflug}].
The corresponding value of $c_\text{CFL}$ is stated for each of the simulations
individually. For the numerical flux, the \ausm scheme (\cref{sec:ausm}) is
used with $\fapdiff=0.1$ and $f_a=\num{e-13}$.  Interface values are calculated
from cell averages by applying linear reconstruction without any slope or flux
limiters. The set of reconstructed quantities is either \rhoP or \rhoT,
depending of the \wbing method, and will be stated explicitly for the
respective simulations. The choice of appropriate boundary conditions depends
on the specific setup, hence they are given for each setup individually. All of
our test cases are two-dimensional for computational reasons, although the
methods are equally valid in three spatial dimensions. Qualitatively different
results for the 3D case are not expected.

\subsection{Simple stratified atmospheres}
\label{sec:hystattest}

A useful test problem for the quality of a well-balanced scheme is a
stably stratified atmosphere. A zero-velocity initial condition should be
maintained perfectly, at least down to rounding errors. However,
in a not well-balanced scheme the pressure-gradient force and gravity do not
cancel exactly and the systems ends up in a state with small, but nonzero,
acceleration. Depending on the details of the numerical flux, this acceleration
may prevent the simulation of flows at very low Mach numbers or it may
cause unphysical convection in formally stable stratifications.
In any case, keeping a hydrostatic stratification stable is a necessary, but
not a sufficient condition for a well-balanced scheme to be useful for the
simulation of stratified atmospheres.

We start with the one-dimensional (1D), isothermal test case of
\citet{kaeppeli2016a} on the domain $[0,\SI{2}{cm}]$. Gravity is points into
negative $x$-direction and the gravitational potential is given by
\begin{equation}
  \phi = s_0 x,
  \label{eq:hystat_phi_linear}
\end{equation}
with a steepness~$s_0$ of \SI{1}{cm.s^{-2}}. The \emph{constant} temperature profile is convectively
stable for any equation of state with a positive value of $\delta$
[\cref{eq:delta_eos}]. The density and pressure profiles fulfilling
\cref{eq:hystat} are given by,
\begin{equation}
  \rho = \rho_0 \exp\left( - \frac{\rho_0}{p_0} \phi \right), \quad
  p   = p_0 \exp\left( - \frac{\rho_0}{p_0} \phi \right).
  \label{eq:isothermal-atmosphere}
\end{equation}
This expression holds for any form of the gravitational potential, not just
\cref{eq:hystat_phi_linear}. The reference density and pressure are set to
$\rho_0 = \SI{1}{g.cm{^{-3}}}$ and $p_0 = \SI{1}{Ba}$, respectively. We use
Dirichlet boundary conditions, which are initialized with the hydrostatic
profile and then left unchanged throughout the simulation. The equation of state
is that of an ideal gas with an adiabatic exponent $\gamma=5/3$.

\begin{figure}
  \includegraphics[width=\columnwidth]{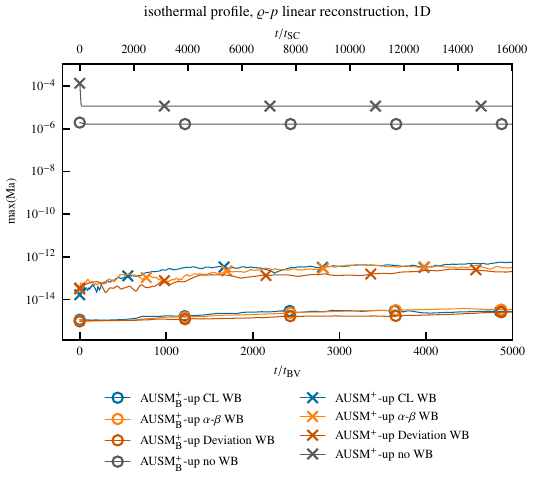}
  \caption{\label{fig:hystattest-isothermal-1d}Time evolution of the maximum
  Mach number in a 1D atmosphere with an isothermal temperature profile
  [\cref{eq:isothermal-atmosphere}] and a linear gravitational potential
  [\cref{eq:hystat_phi_linear}]. The colors indicate different
  well-balancing~(WB) methods, the markers different flux functions. CL stands
  for \leroux. Time is given in units of the \brunt time~$\tBV=\SI{9.93}{s}$
  [\cref{eq:brunt_time}] and sound-crossing time~$\tSC=\SI{3.10}{s}$
  [\cref{eq:sound_crossing}]. The curves have been slightly smoothed for better
  visibility.}
\end{figure}

\Cref{fig:hystattest-isothermal-1d} shows the time evolution of the maximum Mach
number for this isothermal configuration in 1D simulations with 64 grid cells
using a variety of \wbing methods and two different flux functions. The
simulations use linear reconstruction in \rhoP variables, $c_\text{CFL}=0.9$
for the time step size, and they were run until $t=5000\,\tBV$. The figure shows
that runs without \wbing immediately reach Mach numbers between \num{e-6} and
\num{e-5}, while all of the tested \wbing methods manage to keep the Mach number
below \num{e-12}. The choice of flux function does not qualitatively affect the
results here, in particular it does not play a role if we use a low Mach number
flux, such as \ausmup, or a standard flux, such as \ausmp (see \cref{sec:ausm}).

\begin{figure}
  \includegraphics[width=\columnwidth]{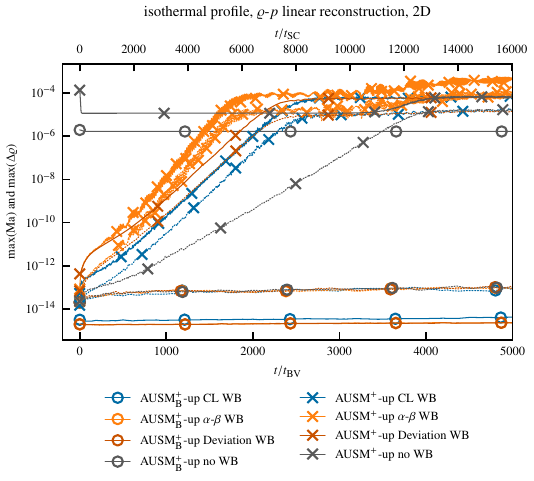}
  \caption{\label{fig:hystattest-isothermal-2d}Same as \cref{fig:hystattest-isothermal-1d}, but for a 2D atmosphere. The solid lines represent the maximum Mach numbers on the grid and the dotted lines the horizontal density fluctuations according to \cref{eq:delta-rho}.  The curves have been slightly smoothed for better visibility.}
\end{figure}

Certain phenomena, such as convection, are inherently multidimensional and
cannot develop in 1D geometry. Therefore we repeat this test using
a two-dimensional (2D) atmosphere. For simplicity we keep gravity pointing in
the negative $x$-direction. The horizontal boundaries are periodic.
\Cref{fig:hystattest-isothermal-2d} shows the evolution of the maximum Mach
number and horizontal density fluctuation~$\Delta\rho$ over time. The latter is
defined as
\begin{equation}
  \Delta \rho = \rho - \langle \rho \rangle_y,
  \label{eq:delta-rho}
\end{equation}
with $\langle \cdot \rangle_y$ denoting an average over the $y$ direction.

Similar to the 1D case, we see that the simulations without \wbing immediately
reach Mach numbers around \num{e-5}. All simulations using any \wbd method
start from very low Mach numbers ($\approx \num{e-13}$), but the ones using the
low Mach number \ausmpup flux  show an exponential growth over the next few
thousand $\tBV$. The growth of the Mach number is linked to that of $\Delta
\rho$. This growth even affects the Mach numbers in the simulation without
\wbing, where the Mach number increases further after about $\num{4000}\,\tBV$.
The \wbd simulations using the \ausmp flux do not show this behavior and
retain the very low Mach numbers. We found that other low Mach number fluxes,
such as the one by \citet{miczek2015a} and \citet{li2008a}, show a growth
similar to \ausmpup, although the rate varies between the different schemes.
These spurious velocities are likely due to pressure--velocity decoupling,
which is a common issue with compressible low Mach number methods. In
combination with a gravity source term this can lead to an instability very
similar to convection, but in stable stratifications. The reason is that
pressure does not immediately return to its horizontal equilibrium. A common
way to partially alleviate this effect is to introduce a form of pressure
diffusion, such as the one suggested by \citet{edwards1998a}, which is also
used in the \ausmpup solver.

While the standard flux seems to perform better in this static test case, it is
ultimately not suited for the simulation of dynamic phenomena at low Mach
numbers, such as convection or waves, due to its high numerical dissipation.

\begin{figure}
  \includegraphics[width=\columnwidth]{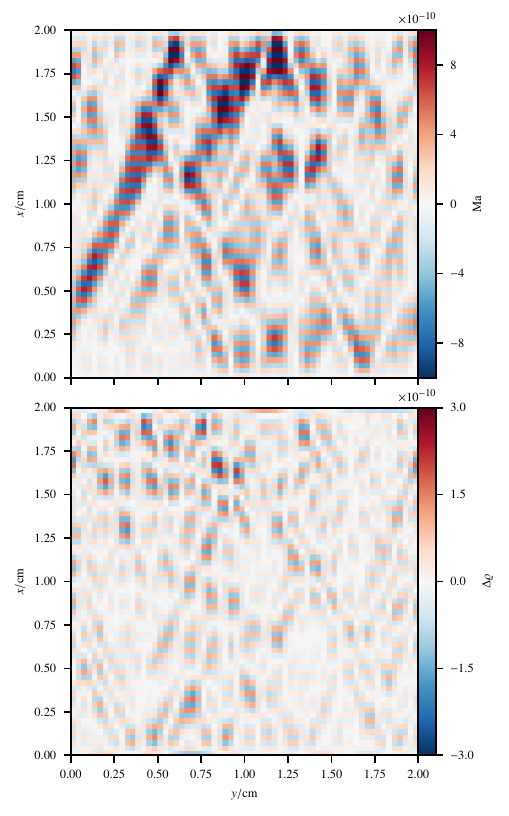}
  \caption{\label{fig:hystattest-mach-hrhofluc}Mach number (\emph{top panel}) and horizontal density fluctuation~$\Delta \rho$ (\emph{bottom panel}) of the 2D isothermal simulation using the \ausmpup flux and deviation \wbing at time $t=\num{1000}\,\tBV$. Gravity is directed in negative $x$-direction.}
\end{figure}

\Cref{fig:hystattest-mach-hrhofluc} shows the typical pattern of the
exponentially growing perturbation in both Mach number and $\Delta \rho$. In
both quantities we see a resolved pattern in the horizontal direction, but a
grid-level oscillation in the vertical direction. One hypothesis for this
behavior is that it is due to unresolved internal gravity waves in combination
with pressure--velocity decoupling, see \cref{subsec:convection}.

We run the same tests for polytropic stratifications. Here the profiles of density, pressure, and temperature are
\begin{equation}
  \rho = \rho_0 \theta^{\frac{1}{\nu - 1}},\quad
  p = p_0 \theta^{\frac{\nu}{\nu - 1}},\quad
  T = \frac{p_0 \mu}{\rho_0 R} \theta = T_0 \theta,
  \label{eq:hystat-polytropic}
\end{equation}
with
\begin{equation}
  \theta = 1 - \frac{\nu - 1}{\nu} \frac{\rho_0}{p_0} \phi.
\end{equation}
We set $\mu=\SI{1}{g.mol^{-1}}$.
The polytropic index~$\nu$ determines the slopes of the profiles. If $\nu$ is
less than the adiabatic exponent~$\gamma$, the atmosphere is stable. If
$\nu>\gamma$, it is unstable. If both are equal the atmosphere is isentropic and
therefore marginally stable.

\begin{figure}
  \includegraphics[width=\columnwidth]{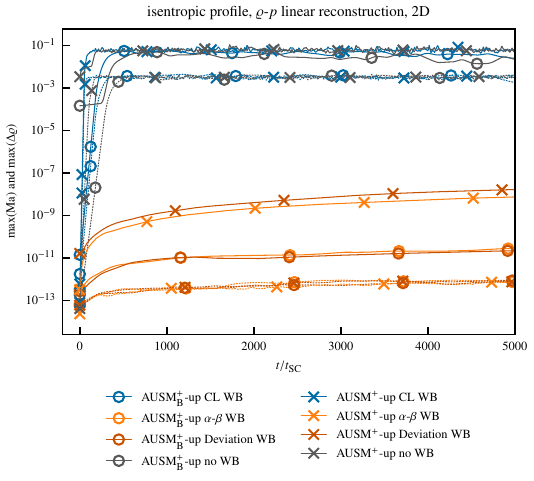}
  \caption{\label{fig:hystattest-isentropic-2d}Same as
  \cref{fig:hystattest-isothermal-2d}, but for an isentropic stratification. The
  solid lines represent the maximum Mach numbers on the grid and dotted lines the
  horizontal density fluctuations according to \cref{eq:delta-rho}. Time is
  given in units of the sound-crossing time~$\tSC=\SI{4.28}{s}$.  The curves
  have been slightly smoothed for better visibility.}
\end{figure}

\Cref{fig:hystattest-isentropic-2d} shows the maximum Mach number and $\Delta
\rho$ for an isentropic atmosphere. Here the \brunt time~\tBV is not well
defined because $N=0$. Instead we use the sound-crossing
time~$\tSC=\SI{4.28}{s}$ for reference. \alphabeta and deviation \wbing
stay at very low Mach numbers below \num{e-7}, independently of the choice of
flux function. This result suggests that the issue with the exponential growth
of perturbations in combination with low Mach number flux functions is more
severe in more stable stratifications.
In contrast to the isothermal test we see that \leroux \wbing behaves quite
similarly to the not \wbd case. Both quickly reach Mach numbers of about
\num{e-1} with flow patterns that resemble 2D convection. It is likely that
this marginally unstable stratification experiences the growth of convection
due to the less than ideal \wbing properties of the \leroux method.

\begin{figure}
  \includegraphics[width=\columnwidth]{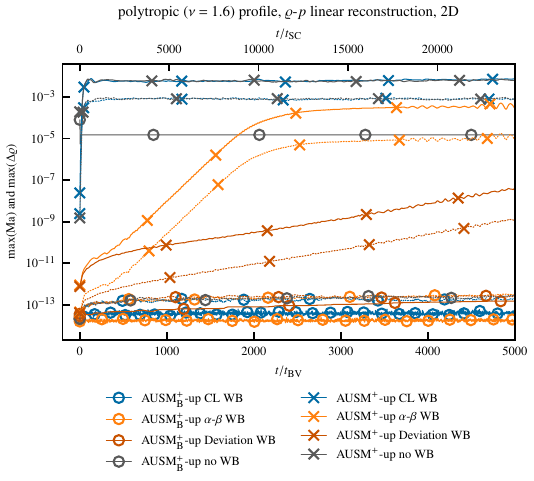}
  \caption{\label{fig:hystattest-polytropic-nu-1_6-2d}Same as
  \cref{fig:hystattest-isothermal-2d}, but for a polytropic stratification with
  $\nu=1.6$. The adiabatic exponent is $\gamma=5/3$. The solid lines represent
  the maximum Mach numbers on the grid and dotted lines the horizontal density
  fluctuations according to \cref{eq:delta-rho}. Time is given in units of \brunt
  time~$\tBV=\SI{20.1}{s}$ and sound-crossing time~$\tSC=\SI{4.13}{s}$. The
curves have been slightly smoothed for better visibility.}
\end{figure}

As a last test we run a polytropic stratification with
\mbox{$\nu=1.6<\gamma=5/3$}, which is stably stratified, but less so than the
isothermal case. The relevant timescales here are $\tBV=\SI{20.1}{s}$ and
$\tSC=\SI{4.13}{s}$. The results are shown in
\cref{fig:hystattest-polytropic-nu-1_6-2d}. Here we see that the cases without
\wbing and with \leroux \wbing fail to preserve the hydrostatic equilibrium in
combination with the low Mach number flux. For the other two \wbing methods
there is a much clearer difference in the growth rate of the perturbation with
deviation \wbing growing significantly more slowly, reaching only Mach numbers
of \num{e-7} after $\num{5000}\,\tBV$. Any of the \wbing methods manages to
keep the Mach numbers below \num{e-11} when combined with the \ausmp flux,
which does not have low Mach number properties. This is in agreement with the
findings in the isothermal case.

\Cref{appx:hystattest} shows the same isentropic and polytropic tests as in
\cref{fig:hystattest-isentropic-2d,fig:hystattest-polytropic-nu-1_6-2d}, but in
the 1D case. In contrast to the 2D cases, the Mach numbers stay at low values
for the runs using \wbing also when using the \ausmpup solver. This is consistent
with the hypothesis that these velocities are a form of unphysical convection
caused by pressure--velocity decoupling, which is obviously not possible in
only one spatial dimension.

The spurious growth we found in combination with low-Mach-number fluxes is
likely not a problem particular to the \wbd methods we presented. We considered
very long timescales of thousands of sound-crossing and \brunt times. This is
much longer than the timescales typically used to test \wbing methods.
\citet{kaeppeli2016a}, for example, ran their hydrostatic setup only for
$2\,\tSC$.\footnote{\citet{kaeppeli2016a} show another test of a 3D simulation
that covers several convective turnover times and contains a stable layer. The
simulation shows strong, unphysical flows in the stable layer when run without
a \wbd method, but these disappear when their \wbd method is used. Considering
they did not use a low Mach number method, this is consistent with our results.}
While this would definitely show any major issues with the basic \wbd property
of the scheme, it does not reveal the long-term growth of instabilities in the
stable region. This is something to bear in mind when applying such methods to
partly convectively stable configurations, such as stars. Whether the described
phenomenon is an issue for real-world applications depends on many factors,
such as how well the stratification is resolved and what timescales are of
relevance. In particular for applications with large stable regions and long
simulation times, such as in asteroseismological hydrodynamics simulations,
this has to be carefully considered. The most promising method in this test is
the deviation \wbing method in combination with the \ausmpup flux.

\subsection{Hot bubble}
\label{sec:hotbubble}
Convection in the stellar interior is usually slow and almost adiabatic. A
typical convection zone is nearly isentropic, although the stratification in the
star's gravitational field can span orders of magnitude in pressure and density.
The buoyant acceleration of a fluid parcel is given by its entropy fluctuation
with respect to the mean entropy at any given radius. Entropy fluctuations are
constantly created by sources of heating or cooling. However, once the fluid
parcel has left the heating or cooling layer and travels through the rest of the
convection zone, the parcel's entropy must be preserved except for those parts
that have mixed with their surroundings. If the flow is slow, all entropy
fluctuations inside the convection zone are also small and it becomes
numerically challenging to preserve their exact values when density and pressure
change by large factors as the fluctuations are advected along the gravity
vector.

We test the numerical schemes' entropy-preservation properties under the
conditions described above by simulating the buoyant rise of a ``bubble'' with
an adjustable initial entropy fluctuation embedded in a layer of constant
entropy. The layer is strongly stratified in pressure and density due to the
presence of gravity. We call this setup ``hot bubble'', because we use positive
initial entropy fluctuations. Negative initial entropy fluctuations would make
the bubble fall, but everything else would work in the same way. Our setup is
similar to that used by \citet{almgren2006a} to test their MAESTRO code,
although their equation of state and stratification differ from our setup. We
also test our methods down to much lower Mach numbers.

We construct the stratification in a two-dimensional box $10^6$\,cm wide and
spanning the vertical range from $y = 0$ to $y = y_\mathrm{max} =1.5 \times
10^6$\,cm. To avoid any influence of boundary conditions, we use periodic
boundaries and a periodic profile of gravitational acceleration
\begin{equation} g_y = g_0 \sin(k_y y), \end{equation}
where $g_0$ is a constant to be specified later and $k_y = 2\pi /
y_\mathrm{max}$. The box is filled with an ideal gas with the ratio of specific
heats $\gamma = 5/3$ and mean molecular weight $\mu = \SI{1}{g.mol^{-1}}$. At $y
= 0$, we set the pressure to $p_0 = 10^6$\,Ba and the temperature to $T_0 =
300$\,K. The stratification is isentropic, that is
\begin{equation} \rho(y) = \left( \frac{p(y)}{A} \right)^{1/\gamma},
\end{equation}
where $A = A_0 = \mathrm{const.}$ everywhere outside of the bubble. Inside the
bubble, we perturb the entropy via
\begin{equation} A = A_0\left[1 + \left( \frac{\Delta A}{A} \right)_{t=0}
\cos\left( \frac{\pi}{2} \frac{r}{r_0} \right)^2 \right], \end{equation}
where $(\Delta A/A)_{t=0}$ is the bubble's amplitude and \mbox{$r = \left[ (x -
x_0)^2 + (y - y_0)^2 \right]^{1/2}$} is the distance from the bubble's center
with $x_0 = 5 \times 10^5$\,cm and $y_0 = 1.875 \times 10^5$\,cm. The bubble has
a radius of $r_0 = 1.25 \times 10^5$\,cm. We do not perturb the hydrostatic
pressure stratification, which is given by the expression
\begin{equation} p(y) = \left\{ p_0^{1 - \frac{1}{\gamma}} + \left(1 -
\frac{1}{\gamma} \right) \frac{g_0}{A_0^{\frac{1}{\gamma}} k_y} \left[ 1 -
\cos(k_y y) \right] \right\}^{\frac{\gamma}{\gamma - 1}}. \end{equation}
The amplitude $g_0$ of the gravity profile sets the ratio of the maximum to the
minimum pressure in the periodic pressure profile. To make the problem
numerically challenging, we use a pressure ratio of $100$, which is achieved
with \mbox{$g_0 = -1.09904373 \times 10^5$\,cm\,s$^{-2}$}. This stratification
is stronger than that in the convective core of a typical massive main-sequence
star, in which the pressure changes by a factor of a few. On the other hand, the
relative pressure drop from the bottom of the solar convective envelope to the
photosphere is ${\approx}10^9$.

\begin{figure*}
  \includegraphics[width=\textwidth]{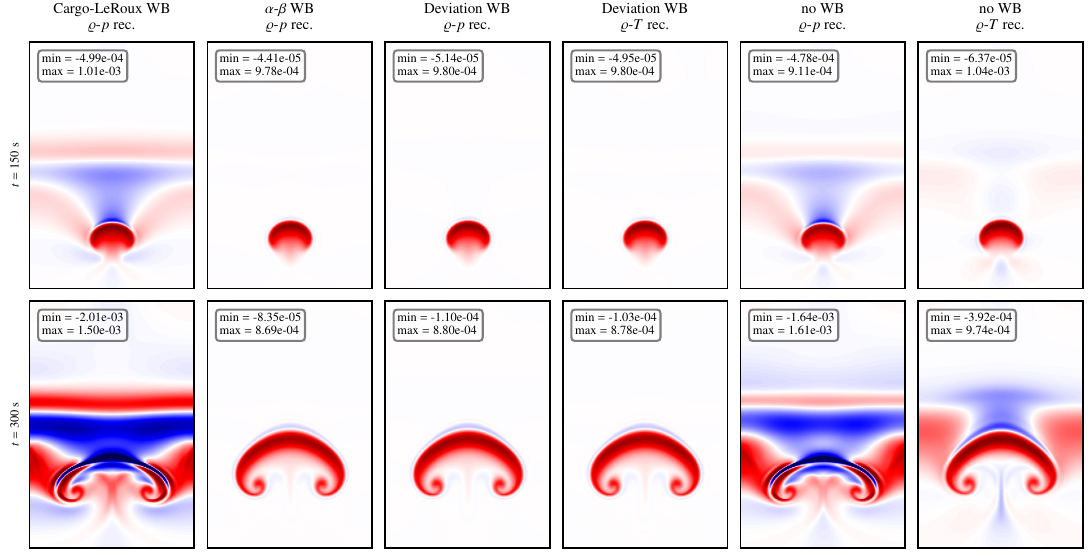}
  \caption{Time evolution (top to bottom) of the hot-bubble problem when solved
      using different well-balancing and reconstruction schemes (left to right)
      on a $128 \times 192$ grid. In all of the cases, entropy fluctuations
      $\Delta A / A$ are shown on the same color scale ranging from $-10^{-3}$
      (dark blue) through $0$ (white) to $10^{-3}$ (dark red). The minimum and
      maximum values of $\Delta A / A$ in the whole simulation box are given in
      each panel's inset. The amplitude of the initial entropy perturbation is
  $(\Delta A/A)_{t=0} = 10^{-3}$.}
  \label{fig:hotbubble-time-dependence}
\end{figure*}

We start with a moderate initial entropy perturbation of $(\Delta A/A)_{t=0} =
10^{-3}$, which makes the bubble rise at moderately low Mach numbers of a few
times $10^{-2}$. This allows us to perform simulations with all three \wbing
methods as well as simulations without any \wbing at modest grid resolution of
$128 \times 192$, see Fig.~\ref{fig:hotbubble-time-dependence}. We run this
series of simulations with fixed time steps of $0.2$\,s. With the exception of
the \leroux and \alphabeta methods, which require \rhoP reconstruction, we test
both \rhoP and \rhoT reconstruction. The central, most buoyant, part of the
bubble accelerates fastest. The bubble gets deformed into a mushroom-like shape
with two trailing vortices and it expands as it rises into layers of lower
pressure. Ideally, the initially positive entropy fluctuations $\Delta A/A$
should mix with the isentropic ($\Delta A/A = 0$) background stratification,
creating smaller but still positive entropy fluctuations. The entropy
fluctuations may locally increase a bit as kinetic energy is slowly dissipated
into heat, but there is no physical way for them to become negative. Any
negative entropy fluctuations in the numerical solution result from numerical
errors.

Figure~\ref{fig:hotbubble-time-dependence} shows that both the absence of
well-balancing and the \leroux method generate large areas of negative entropy
fluctuations comparable to or even larger in absolute value than the bubble's
initial amplitude. Large-scale positive entropy fluctuations also occur far from
the bubble and they clearly do not result from hydrodynamic mixing. They rather
seem to be caused by entropy nonconservation as the bubble pushes the
surrounding stratification upwards and downwards. In the no-well-balancing case,
errors in $\Delta A/A$ are smaller when \rhoT reconstruction is employed  as
compared with \rhoP reconstruction. This may be due to the fact that the
pressure changes by a factor of $100$ in the computational domain whereas the
temperature only changes by a factor of $6.3$. In any case, \alphabeta and
deviation \wbing clearly provide far superior results with only a mild and
highly localized undershoot in $\Delta A/A$ above the bubble. With the deviation
method, this success is independent of the choice of reconstruction.

\begin{figure*}
  \includegraphics[width=\textwidth]{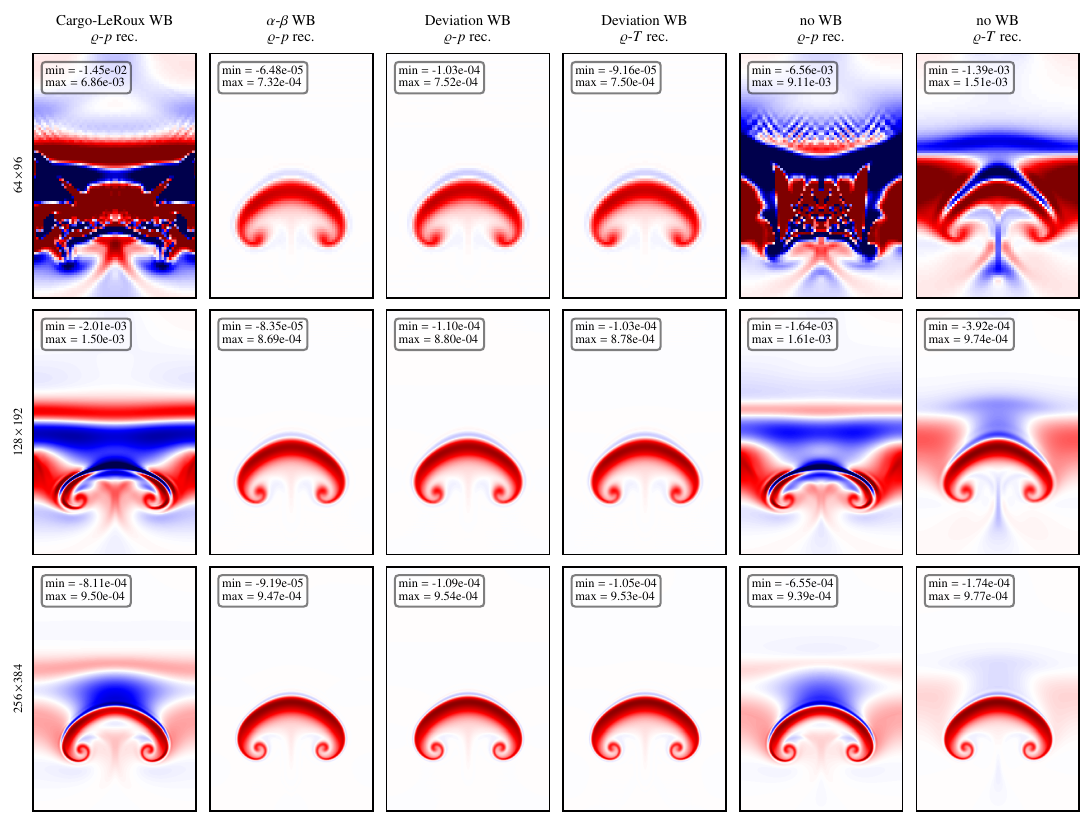}
  \caption{Same as Fig.~\ref{fig:hotbubble-time-dependence}, but showing the
  solutions' resolution dependence (top to bottom) at $t = 300$\,s.}
  \label{fig:hotbubble-resolution-dependence}
\end{figure*}

All of the methods tested converge upon grid refinement, although not in the
same way, see the series of runs shown in
Fig.~\ref{fig:hotbubble-resolution-dependence}. In this series, we keep the
$0.2$\,s time steps for all runs except those on the finest ($256 \times 384$)
grid, for which we use $0.1$\,s. When there is no \wbing or the \leroux method
is used, the amplitude of the large-scale entropy-conservation errors around
the bubble decreases upon grid refinement and the bubble's shape slowly
approaches that obtained with the \alphabeta and deviation methods. The errors
are still substantial even on the finest ($256 \times 384$) grid tested. The
slight entropy undershoot produced by the \alphabeta and deviation methods does
not decrease in amplitude but it does decrease in spatial extent as the grid is
refined. Surprisingly, we do not observe any significant entropy fluctuations
far from the bubble even on the coarsest grid ($64 \times 96$) with these two
methods.

As the bubble's initial amplitude $(\Delta A/A)_{t=0}$ is decreased, the typical
Mach number in the flow field decreases as
\begin{equation}
    \mathrm{Ma} \propto \left( \frac{\Delta A}{A} \right)_{t=0}^{1/2}.
    \label{eq:hotbubble-Ma-scaling}
\end{equation}
This scaling results from the fact that the bubble's acceleration is
proportional to $\Delta \rho/\rho \propto \Delta A/A$ and the velocity an object
in uniformly accelerated motion reaches over a fixed distance (i.e., until the
bubble has reached the same evolutionary stage) is proportional to the square
root of the acceleration. Although the bubble's acceleration is not constant,
the bubble always evolves in the same way, just slower, when its initial
amplitude is decreased and the scaling still holds.

\begin{figure}
  \includegraphics[width=\columnwidth]{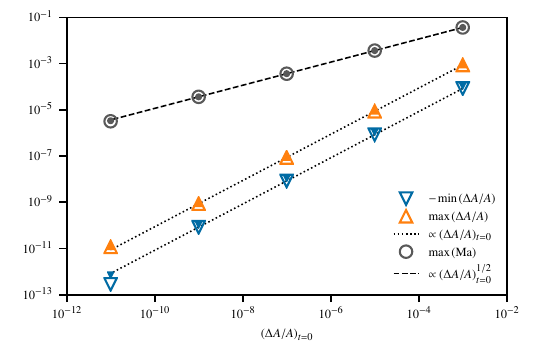}
  \caption{Dependence of the maximum Mach number $\mathrm{Ma}$ and minimum and
      maximum entropy fluctuations $\Delta A/A$ on the bubble's initial
      amplitude $(\Delta A/A)_{t=0}$. All measurements are taken when the bubble
      has evolved to a stage comparable to that shown in the bottom row of
      Fig.~\ref{fig:hotbubble-time-dependence}. Open and filled symbols
      correspond to runs with \alphabeta and deviation \wbing, respectively. The
      lines show that the Mach number scales according to
  Eq.~\eqref{eq:hotbubble-Ma-scaling} and the minimum and maximum entropy
  fluctuations scale in proportion to the bubble's initial amplitude.}
  \label{fig:hotbubble-amplitude-scalings}
\end{figure}

The scaling is also demonstrated in Fig.~\ref{fig:hotbubble-amplitude-scalings},
which shows a series of runs performed on a $128 \times 192$ grid with $(\Delta
A/A)_{t=0}$ ranging from $10^{-3}$ down to $10^{-11}$. As $(\Delta A/A)_{t=0}$
decreases, we increase time steps in this order: $0.2$\,s, $1$\,s, $10$\,s,
$25$\,s, $25$\,s. The solver's convergence worsens as all fluctuations become
smaller, limiting the maximum time step size. We only include the \alphabeta
and deviation \wbing methods in this experiment, because entropy-conservation
errors quickly dominate the solution when the initial amplitude is decreased
and the \leroux or no \wbing method is used. Because the Mach number is
expected to scale according to Eq.~\ref{eq:hotbubble-Ma-scaling}, we scale the
time when the simulation is stopped with $(\Delta A/A)_{t=0}^{-1/2}$. This way,
we compare the results when the bubble has reached approximately the same
height and evolutionary stage as the previously discussed case with $(\Delta
A/A)_{t=0} = 10^{-3}$ at $t = 300$\,s
(Fig.~\ref{fig:hotbubble-time-dependence}). Both the \alphabeta and deviation
\wbing methods provide essentially the same solutions, reproducing the expected
Mach-number scaling down to $\mathrm{Ma} \sim 10^{-6}$. The amplitude of the
entropy undershoot above the bubble is $24\%$ lower when the \alphabeta method
is used as compared with the deviation method. The most extreme run with
$(\Delta A/A)_{t=0} = 10^{-11}$ reaches the limits of our current implementation
and we can see numerical noise developing in the stratification, see
Figs.~\ref{fig:hotbubble-amplitude-dependence} and
\ref{fig:hotbubble-amplitude-dependence-Ma} in \cref{appx:hotbubble}.
Figure~\ref{fig:hotbubble-amplitude-scalings} also shows that the minimum and
maximum entropy fluctuations in the evolved flow scale in proportion to the
initial amplitude of the bubble.

\subsection{Simple convection setup}
\label{subsec:convection}

The previous section demonstrated the capabilities of the \alphabeta and
deviation \wbing schemes to evolve the rise of a bubble with an entropy excess
in an isentropic stratification at low Mach numbers. This can be interpreted as
a test for the fundamental mechanism of convection. The purpose of this section
is to also assess the benefit from \wbing techniques for fully developed
convection in a realistic stellar scenario.

The prototype of a convective region in stellar interiors includes a steady
heating source in the form of nuclear burning that injects energy over a long
period in time. If radiative transport of energy is not efficient enough, the
temperature gradient will steepen until it reaches the adiabatic temperature
gradient. According to \cref{eq:brunt_nabla}, this region will become unstable
and convection will set in. As convection is very efficient in transporting
energy, a common assumption is that the stratification settles to an nearly
adiabatic temperature gradient in the steady state.

In stars, a convective region is typically adjacent to convectively stable
regions. The mixing processes across the interfaces of convectively stable and
unstable regions have a profound impact on stellar evolution, yet it is
particularly difficult to parametrize these processes in one-dimensional
stellar evolution codes. A common strategy for improving the current 1D
description is to investigate the dynamics at the interfaces of convection
zones by means of multidimensional hydrodynamical simulations
\citep{jones2017a, cristini2019a, pratt2020a, higl2020a, horst2021a}.

The Mach number of convection deep in the stellar interior is estimated to be
on the order of \num{e-4} in early stages of stellar evolution.  Accurate
modeling of the early phases is, however, crucial as it determines the whole
subsequent evolution of the star and inaccuracies will propagate to later
stages. Thus, numerical experiments that address convection in stellar
interiors rely on schemes that accurately maintain hydrostatic stratifications
and that are able to follow convection at low Mach numbers for a sufficient
amount of time.

The initial stratification for the test series presented in this section
consists of a convective region with a stable layer on top. One possibility to
set up this configuration would be to use realistic initial conditions from a
1D stellar evolution code. However, such 1D profiles usually require some extra
treatment before they can be used for hydrodynamical simulations, for example a
smoothing of sharp composition gradients or to properly impose a flat entropy
profile in the convection zone. Instead, we use analytical initial conditions
to be able to test the numerical methods under well-defined but realistic
conditions. This way, any numerical artifacts that may arise can solely be
attributed to the methods applied rather than to inadequate initial conditions.

To construct the initial hydrostatic stratification, we follow the procedure
described by \citet{edelmann2017a}. It imposes the profile of the
superadiabaticity $\supad = \nabla - \nabla_\text{ad}$ while integrating the
equation of hydrostatic equilibrium [\cref{eq:hystat}]. According to
\cref{eq:brunt_nabla}, $\supad$ determines the sign of the \brunt frequency. It
is therefore possible to precisely control which regions are convectively
stable ($\supad<0$) and unstable ($\supad>0$) as well as the respective
transitions between these regions. A marginally stable stratification is
imposed inside the convection zone by setting $\supad_{\cz} = 0$.  In the
stable region, we impose $\supad_{\stab} = -\nabla_\text{ad}$, which
corresponds to an isothermal stratification.

To connect the two regions, we use a sinusoidal transition, which ensures that
the transition between the two $\supad$-values is well-defined and can be
resolved numerically. The profile with a transition between the convection
and stable zones then takes the form
\begin{align}
  \supad(y) &= \supad_{\cz} \notag\\
  &+ \frac{1}{2}\left[1 +
  \sin\left(\frac{\pi}{2}\,\eta(y, K, y_{\text{\cz,\stab}})\right)\right]\left(\supad_{\stab}-\supad_{\cz}\right), \label{eq:sintrans}
\end{align}
with
\begin{align}
  \eta(y, K, y_{i}) = \left\{
  \begin{aligned}
    -1 & \quad\text{ if } K(y-y_{i}) < -1,\\
    1 & \quad\text{ if } K(y-y_{i}) > 1,\\
    K(y-y_{i}) & \quad \text{ otherwise,}
  \end{aligned}
  \right. \qquad
\end{align}
where the constant $K$ determines the steepness of the transition. It is set
such that the transition is resolved by at least \num{20} grid cells. The
coordinate \ycs denotes the middle of the transition starting at $\ycs-1/K$ and
ending at $\ycs+1/K$. The value of \ycs is given in \cref{tab:convparams}. The
computational domain spans $2\ycs$ in the horizontal direction.
The profiles of temperature $T$, pressure $p$ and density $\rho$ then follow
from integrating the equation of hydrostatic equilibrium \cref{eq:hystat} as
described by \citet{edelmann2017a} with a spatially constant gravitational
acceleration of $\left|\g\right| =
\SI{6.6e4}{\centi\meter\per\square\second}$. The initial values required for
the integration are listed in \cref{tab:convparams}. They are representative
of the conditions expected in the convective core of a \SI{25}{\msun} main
sequence star with a mean atomic weight of $\overline{A}=1.3$ and a mean
charge of $\overline{Z}=1.1$.

\begin{figure}
  \includegraphics{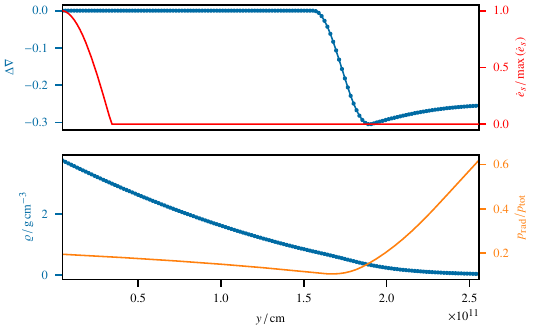}
  \caption{Initial stratification of the convection setup. The red curve
  illustrates the position and shape of the energy injection which has nonzero
  values only at the bottom of the convection zone. Its actual amplitude is set
  for the different simulations individually. Dots denote the positions of cell
  centers on a grid with $144$ vertical cells, the lowest resolution used in
  the \slh simulations presented in this section.}
  \label{fig:convbox_init}
\end{figure}

\Cref{fig:convbox_init} shows the resulting profiles of \supad, $T$, and
$\rho$ as well as the fact that the transition between the convective and
stable zone is well resolved even on the coarsest \slh grid. Because cores of
massive stars are hot, a considerable fraction of the total pressure is
contributed by the radiation field. The relative importance of radiation
pressure is given in the lower panel of \cref{fig:convbox_init}. It ranges from
roughly \SI{20}{\percent} in the bottom region to about \SI{60}{\percent} in
the top region where the density is low. This test setup is therefore also an
example where the ideal gas \eos is not sufficient to describe the
thermodynamic behavior of the gas and which requires \wbing methods that can
handle general \eos{}.

To trigger convection in the initially marginally stable convection zone, a heat
source is placed at the bottom of the convection zone that continuously injects
energy into the system with a sinusoidal profile peaking at the bottom of the
domain. The heating profile is given by
\begin{align}
  \dot{e}_h(y) = \dote\,a\, \sin\left(\frac{\pi}{2}\left[1+\eta\left(y, K, 0\right)\right]\right)\,\si{erg.s^{-1}.cm^{-3}},
  \label{eq:convbox_driving}
\end{align}
with
\begin{align}
  a = \frac{4\sin(\pi K \Delta y/4)}{\pi K \Delta y},\,
  K = \frac{1}{0.2\,y_{\cz,\stab}},
\end{align}
where $\Delta y$ denotes the grid spacing in the vertical direction. As
introduced here, $\dot{e}_0$ is dimensionless. The factor $a$ ensures that the
total heating rate is independent of grid resolution.
\begin{table}
  \centering
  \caption{Parameters of the convection test setup.}
  \label{tab:convparams}
  \begin{tabular}{ll}
    Quantity & Value \\
    \toprule
    $T_0$                                       & \SI{3.7e7}{\kelvin} \\
    $p_0$                                       & \SI{2.4e16}{\gram\per\centi\meter\per\second\squared} \\
    $|\g|$                                      & \SI{6.6e4}{\centi\meter\per\second\squared} \\
    $y_\text{bot},\, y_\text{top}$              & \SI{0}{},\, \SI{2.6e11}{\centi\meter} \\
    \ycs                                        & \SI{0.66}{y_\text{top}} \\
    $x_\text{right} - x_\text{left}$            & \num{2}\,\ycs \\
    $\supad_{\cz},\, \supad_{\stab}$            & $\num{0},\, -\nabla_{\text{ad}}$ \\
    $\overline{A},\, \overline{Z}$              & \num{1.3},\, \num{1.1} \\
    \bottomrule
  \end{tabular}
\end{table}
In a stationary state, the convective velocity is expected to scale with the
heating rate,
\begin{align}
  \mrms \propto \dot{e_0}^{1/3}.
  \label{eq:mascal}
\end{align}
This scaling can be motivated using dimensional arguments \citep[e.g.,
see][]{jones2017a} or the mixing-length theory \citep{kippenhahn2012a} and has
been confirmed in numerical experiments (e.g.,
\citealp{cristini2019a,edelmann2019a,andrassy2020a}).

To test the low-Mach capabilities of the methods that are presented in this
paper, the amplitude of the heating factor $\dote$ is decreased successively
while the root mean square (rms) Mach number measured from the simulation is compared to the
expected scaling \cref{eq:mascal}. The upper limit of $\dote$ is chosen such
that the resulting Mach number is in a regime where also simulations without
\wbing follow the scaling. A deviation from the expected scaling at lower
values of $\dote$ and correspondingly lower Mach number then indicates that
numerical errors have become significant and the method has reached its limit
of applicability.
The scaling test is performed for all available \wbing methods and also in the
absence of \wbing. The domain is discretized by $\num{72}\times\num{144}$
cells. This rather coarse resolution is chosen to assess the ability of the
\wbing method to balance hydrostatic stratifications even with a moderate
number of cells. While any consistent method will ultimately be able to follow
low-Mach-number flows given a sufficiently fine grid, this is computationally
not feasible, especially in 3D.  The simulations use periodic
boundary conditions in the horizontal direction. At the top and bottom of the
domain solid-wall boundaries are employed that do not allow mass to enter or
leave the domain. For the time stepping, we set $c_\text{CFL}=0.5$. The initial
stratification is perturbed by random density fluctuations at the
\ord{\num{e-14}} level to facilitate the growth of the convective instability.
Additionally, the heating profile \cref{eq:convbox_driving} is modulated by a
sinusoidal function along the horizontal direction to break the initial
horizontal symmetry. The wavelength is one fourth of the horizontal extent and
the amplitude is set to $0.01\, \dot{e}_h(y)$. The modulation is switched off
after the flow has reached a substantial fraction of its final speed.

For all simulations, a grid file is saved every \num{200} time steps. For each
saved grid, the mass-weighted rms Mach number \mrms is calculated. To this
end, only the fixed region from $y_\text{bot}$ to \ycs is considered, although
the position of the boundary between convective and stable zone is dynamical
and may change over time. However, for the simulation presented in this
section, this effect is negligible and a fixed region is chosen for
convenience. For all simulations, \mrms is then averaged over the same time
span in terms of the convective turnover time \si{\tconv}, which we define as
\begin{align}
   \tconv = \frac{2\, \ycs}{v_\text{rms}},
   \label{eq:tconv}
\end{align}
where $v_{\text{rms}}$ is the rms velocity.  This ensures
that the stochastic fluctuations, which have different typical timescales for
different flow speeds, are accounted for in a similar way for all simulations.
Due to the steady heat injection and the small amount of numerical dissipation
in 2D simulations, the value of \tconv slightly decreases over time as
velocities slightly increase. Thus, the average of \tconv depends on the
time interval considered and is not clear how to chose the proper time interval
for the different simulations. Instead, the number of turnover times $N_\tau$
as a function of physical time $t$,
\begin{align}
  N_\tau(t) = \int_{0}^{t}\frac{1}{\tconv(t')}\,\text{d}t',
\end{align}
is used to determine the respective time intervals and the averaging is done in
the time interval $t\in\left[t(N_\tau=5),\,t(N_\tau=10)\right]$.
\begin{figure}
  \centering
  \includegraphics[width=\columnwidth]{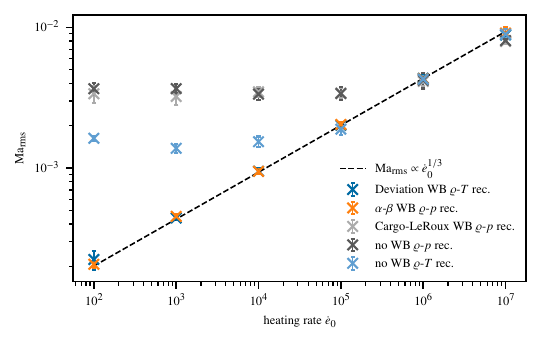}
  \caption{Root mean square Mach number \mrms as a function of the heating rate
  $\dote$. The dashed line represents the expected scaling according to
  \cref{eq:mascal}. All values correspond to time averages over
  $t\in\left[t(N_\tau=5),\,t(N_\tau=10)\right]$ (see text). The vertical
  error bars denote one standard deviation of the temporal average of \mrms.}
  \label{fig:convbox_line_scaling}
\end{figure}
The resulting \mrms as functions of the heating rate are depicted in
\cref{fig:convbox_line_scaling}. For the two highest values of $\dote$, that is
at \mrms around \num{9e-3} and \num{4e-3}, all methods agree and \mrms follows
the scaling given by \cref{eq:mascal}. This is not the case at lower heating
rates. For $\dote\lesssim\num{e5}$, the simulation using the
\leroux\wbing method considerably deviates from the expected scaling by giving
a value of \mrms that does not correlate with the energy input anymore but
stays rather constantly slightly below \num{4e-3}.  Almost identical results
are found when no \wbing is applied in combination with \rhoP as reconstruction
variables. At this point, it seems that \leroux\wbing is not able to improve
the behavior at lower Mach numbers. The results slightly improve if $\rho$-$T$
are used for reconstruction and no \wbing is used. In this case, the Mach
number settles slightly below $\mrms = \num{2e-3}$ for an energy input rate
lower than $\dote=\num{e5}$. The reason for this could be the nonnegligible
contribution of radiation-pressure to total pressure. As $p_\text{rad}\propto
T^4$, a more accurate reconstruction of $T$ could lead to an interface pressure
that is closer to the hydrostatic solution and hence artifacts from imperfect
balancing are reduced.

By definition, \alphabeta\wbing requires \rhoP  to be reconstructed while the
variables can be chosen freely for deviation \wbing.  Hence we have chosen
$\rho$-$T$ for the deviation runs for comparison. However, no major differences
can be seen between \alphabeta and deviation in
\cref{fig:convbox_line_scaling}.  Both \mrms profiles  closely follow the
scaling down to the smallest value of \dote. Due to the hydrostatic solution's
being stored at cell interfaces in these two methods, the particular choice of
variables for reconstruction seem to be less important. In contrast, the
\leroux method reconstructs the potential $q$ at the interfaces using values at
the cell centers, which can introduce an error in the total energy over many
time steps. A possible fix for this would be to store $q$ at the cell
interfaces, however this is currently not implemented in our code.

\begin{figure*} \centering
  \includegraphics[width=\textwidth]{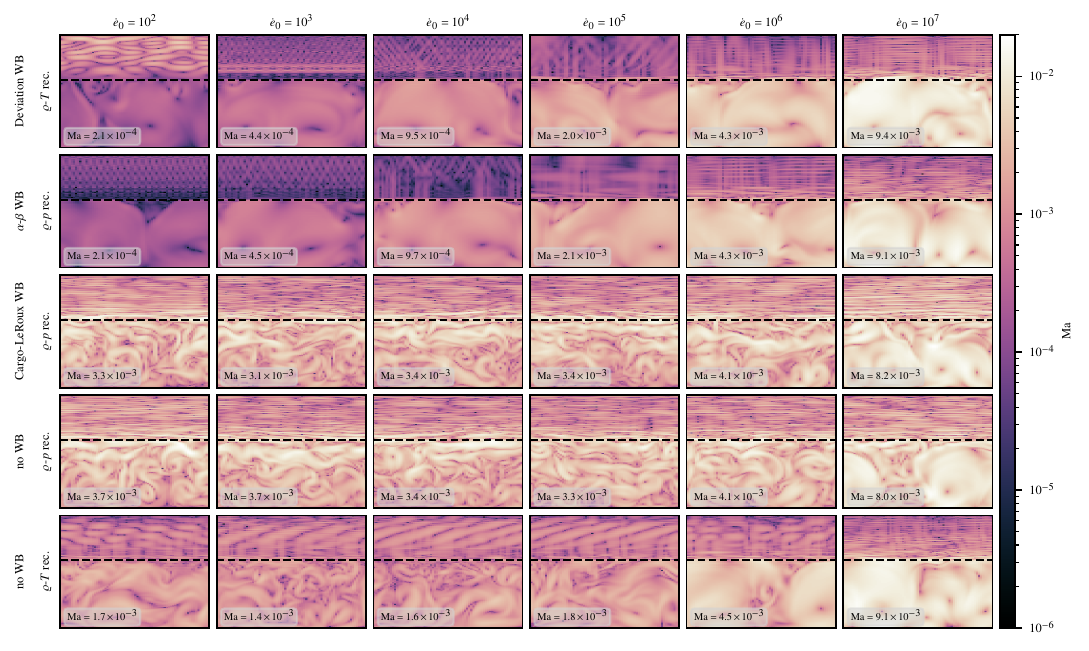}
  \caption{Mach number of the flow for different values of the heating rate
  \dote \textbf{(left to right)} and different \wbing methods
  \textbf{(top to bottom)}. The dashed black lines denote the boundaries of the
  convection zone at $y=y_\text{top}$, see \cref{tab:convparams} for an
  overview of simulation parameters. The insets displays the rms Mach number
  \mrms within the convection zone for the snapshot shown. All snapshots are
  taken at $t(N_\tau=7.5)$.}
  \label{fig:convbox_pcolor_conv}
\end{figure*}

It is not obvious how to assess the accuracy at which hydrostatic
equilibrium is maintained within the convective region. Convection will
inevitably introduce ram pressure, $p_\mathrm{ram}$. Its ratio with thermal
pressure, $p_\mathrm{thermal}$, is expected to scale as
$p_{\mathrm{ram}}/p_\mathrm{thermal} \sim \mrms^2$. This gives an
order-of-magnitude estimate for the expected minimal deviation from the initial
hydrostatic pressure stratification caused by convective motion alone,
independent of the choice of \wbd method. We have verified that the respective
relative deviation from the hydrostatic pressure at $t(N_\tau=7.5)$ scales as
$\mrms^2$ for all sets of simulations. The only exception is the simulation
with deviation \wbing at the lowest heating rate, which may be a result of
sound waves excited by the strong flow field in the stable zone (see
\cref{fig:convbox_pcolor_conv} and the next paragraph). \Cref{tab:hsedev} in
\cref{appx:convbox} lists the ratio exemplarily for the simulations using
\alphabeta\wbing.  There, the relative deviation ranges from about $\num{e-7}$
at the smallest heating rate to roughly $\num{e-4}$ at the largest rate. At
$t=0$ we measure a mean relative deviation of \num{2.7e-8}, which originates
entirely in the discretization error of the pressure gradient. The fact that
the relative deviations from hydrostatic equilibrium are of the same order of
magnitude as the ram pressure implies that they are not an artifact of the
\wbing method, but physically caused by the convective motions.

To also add a qualitative visual verification of the arising convection, the
flow patterns are shown in \cref{fig:convbox_pcolor_conv} in the middle of the
time frame considered. For the highest heating rate, all simulations show the
typical flow morphology of two-dimensional convection: A pair of large eddies
form with a size that is determined by the vertical extent of the convection
zone. At lower heating rates, the flow pattern remains basically the same for
deviation and \alphabeta \wbing, which strengthens our confidence in these
solutions. In contrast, the flow patterns obtained with \leroux \wbing or
without any \wbing are dominated by incoherent, small-scale structures. While
for the ``no WB'' run with $\rho$-$T$ reconstruction, the Mach numbers achieved
are somewhat smaller, the general flow pattern is similar. We assume that these
motions are caused by the imperfect hydrostatic equilibrium.

Convective regions are known to excite internal gravity waves (IGW) in adjacent
stable layers \citep[see e.g.,][]{rogers2013a,edelmann2019a,horst2020a}.
Assuming that IGW are predominantly generated at periods close to the
convective turnover time $\tconv$ \citep[c.f.][]{edelmann2019a}, we can
estimate their vertical wavelength $\lambda_\mathrm{v}$ using the dispersion
relation of IGW in the Boussinesq approximation \citep[see,
e.g.,][]{sutherland2010a},
\begin{align}
   \lambda_\mathrm{v} = \frac{\lambda_\mathrm{h}}{\sqrt{\left[{N
   \tconv}/({2\pi})\right]^2 -1 }},
  \label{eq:igw}
\end{align}
where $\lambda_\mathrm{h}$ is the horizontal wavelength. It follows that the
vertical wavelength decreases with increasing \tconv for $\lambda_\mathrm{h} =
\mathrm{const.}$ and the waves will become unresolved when the heating rate is
decreased on the same computational grid.

\begin{figure}
  \centering
  \includegraphics[width=\columnwidth]{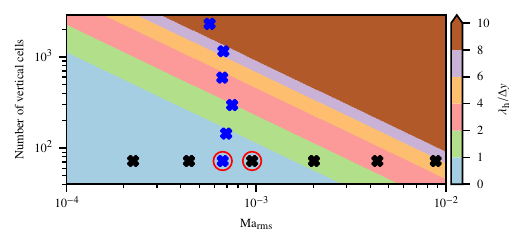}
  \caption{Expected typical vertical wavelength [\cref{eq:igw}] in grid cells
  of internal gravity waves as a function of \mrms in two series of simulations
  with the deviation \wbing method. Black crosses correspond to simulations at
  fixed resolution but increasing heating rate. Blue crosses correspond to
  simulations at a fixed heating rate but different resolution. The two
  encircled data points result from the same simulation but \mrms has been
  determined at a different time.}
  \label{fig:convbox_pcolor_igw_est}
\end{figure}
We estimate the vertical wavelengths according to \cref{eq:igw} exemplarily
for the simulations with deviation \wbing shown in
\cref{fig:convbox_line_scaling,fig:convbox_pcolor_conv}.  To calculate
$\lambda_\mathrm{v}$, the value of the \brunt frequency is taken at the top of
the box domain and $\lambda_\mathrm{h} = 2\,\ycs$, which corresponds to the
maximal horizontal wavelength that fits into the computational domain. The
ratio of $\lambda_\mathrm{v}$ to the vertical grid spacing is shown in
\cref{fig:convbox_pcolor_igw_est} as black crosses. At all but the two
highest heating rates, the vertical wavelength is less than two cells and it
follows from the Nyquist sampling criterion that such waves cannot be
represented on our coarse grid. Indeed, for these runs strange patterns appear
in the stable zone as can be seen in \cref{fig:convbox_pcolor_conv}. A peculiar
pattern is visible in the stable zone for deviation \wbing at the lowest
heating rate. While its origin is not completely understood, we have
verified that the results look similar to the other runs when \rhoP
reconstruction or a higher grid resolution is applied. Our estimate gives
typical wavelengths of two to six cells in the two simulations with the highest
heating rates. Hence, the flow patterns in \cref{fig:convbox_pcolor_conv} for
these two simulation probably include some real internal gravity waves, but
they are still dominated by artifacts.

\begin{figure*}
  \centering
  \includegraphics[width=\textwidth]{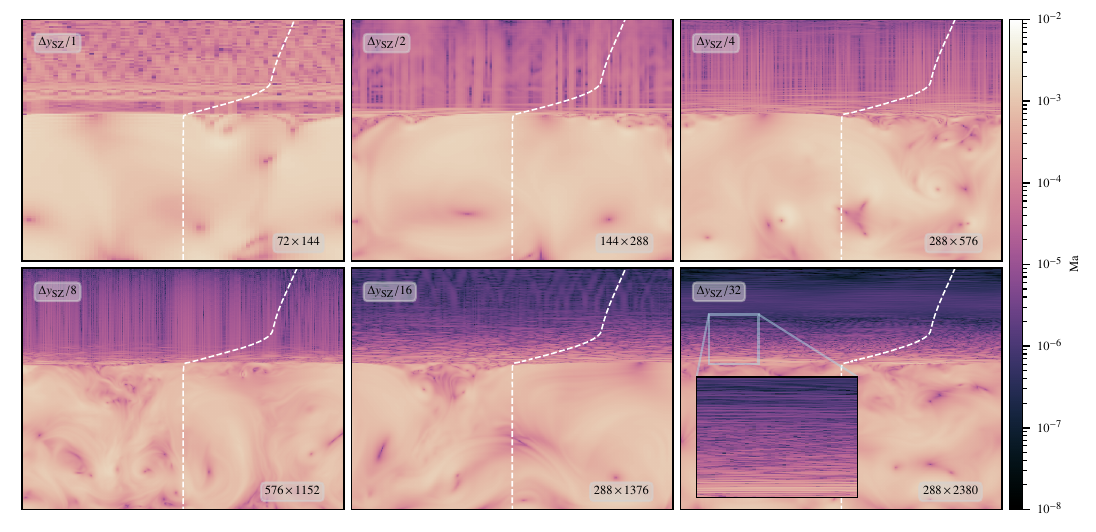}
  \caption{Mach number for a heating rate of $\dote=\num{e4}$ and
  deviation \wbing at different resolutions. The upper left box in each panel
  indicates the change in the vertical resolution relative to the resolution
  used for the Mach number scaling. The lower right box gives the total
  number of cells of the particular simulation. All snapshots are taken as
  soon as  convection has fully developed. The dashed white line illustrates
  the profile of the \brunt frequency as a function of height with arbitrary
  units on the $x$-axis.}
  \label{fig:convbox_pcolor_pat}
\end{figure*}
To confirm that our interpretation is correct, we run another series of
simulations with increasing vertical resolution at $\dote=\num{e4}$.  Deviation
\wbing is used in this experiment. Because of the higher computational costs,
there is not enough data to perform meaningful averages and \mrms is extracted
for a single snapshot as soon as convection has developed in the whole
convective zone. The results are shown as blue crosses in
\cref{fig:convbox_pcolor_igw_est}. Because \mrms is measured at earlier times
compared with the corresponding simulation in the heating series, the \mrms of
the lowest-resolution run does not coincide with the (red circled) black cross
corresponding to the same simulation. We see that the expected typical
vertical wavelength is resolved by more than eight cells on the finest grid.
\cref{fig:convbox_pcolor_pat} shows the flow pattern for the different grid
resolutions. In the center and right panel of the lower row, the resolution in
the convective zone corresponds to the resolution of the upper right panel
($\num{288}\times\num{576}$). Close to the transition to the stable zone, the
vertical spacing is smoothly reduced to $1/32$ of the starting resolution. This
saves computing resources and also illustrates that the patterns do not depend
on the resolution in the convective zone. The transition is shown in
\cref{fig:convbox_vary}.

Our resolution study indicates that with increasing vertical resolution in the
stable zone artifacts are diminished. For the highest resolution fine
wave patterns are observed. As grid resolution increases, nearly horizontal
wave patterns first appear close to the convective boundary, where $N$
gradually increases from zero to relatively large values higher up in the
stable zone.

While our findings can be explained by unresolved IGW, we cannot exclude that
at least to some extent also numerical artifacts contribute to the flows in the
stable zone. Nevertheless, our tests with the simple convective box show that,
as soon as we resolve the IGW sufficiently, artifacts tend to disappear and any
instabilities possibly still present do not visibly dominate the flow.  At the
same time, this illustrates that, depending on the actual stellar profile,
rather high grid resolution is needed to properly resolve the waves.

\subsection{Keplerian disk}
\label{sec:keplerian}
\begin{figure}
\includegraphics[width=\columnwidth]{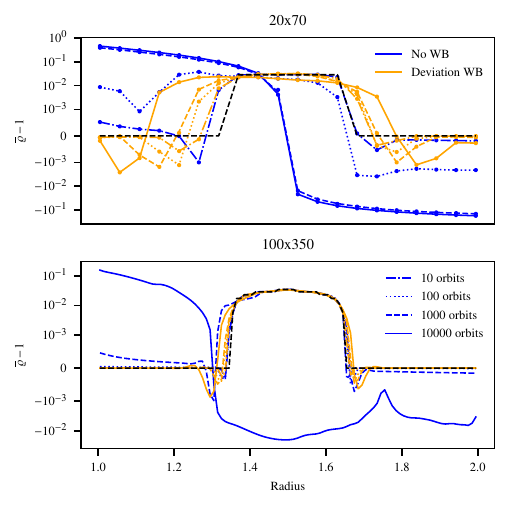}
\caption{\label{p:keplerian_profiles} Time evolution of the angle-averaged
   density profiles $\overline{\rho} - 1$ in the Keplerian disk setup. The top
   panel shows results with a grid resolution of $20 \times 70$ cells, where the
   dots on top of the lines represent
corresponding results computed with implicit time stepping. The bottom panel
shows the same quantity computed on a
much finer grid of $100 \times 350$ cells. The
black dashed lines in each panel show the respective initial density profiles.}
\end{figure}

Some astrophysical problems involve stationary solutions that are not at rest,
for example a rotating star that is partially stabilized by the centrifugal force
\citep[e.g.,][]{tassoul2000a,maeder2009a}. Another case is the Keplerian motion
around a central gravitational mass $m$ in
its gravitational potential $\phi(r) = - \frac{Gm}{r}$. \citet{gaburro2018a}
describe a nondimensional test setup of a circular disk with $\rho_0 = p_0 =
1$ around a massive object. Neglecting its own gravitational field,
such a disk can be stabilized by a flow velocity of
\begin{align}
u(x,y) = & - \sin\left[\mathrm{atan2}\left(y, x\right)\right] \sqrt[]{\frac{Gm}{r(x,y)}} \\
v(x,y) = & \cos\left[\mathrm{atan2}\left(y, x\right)\right] \sqrt[]{\frac{Gm}{r(x,y)}},
\end{align}
where $r = \sqrt[]{x^2 + y^2}$ and $\mathrm{atan2}\left(y, x\right)$ is the
typical shortcut for choosing the quadrant of $\arctan\left(y/x\right)$ correctly. For convenience we set $G=m=1$. We simulate
the Keplerian disk from radius $r=1$ to $2$ on a polar grid with $20$ radial
and $70$ angular cells as well as on a finer grid with $100$ radial and $350$ angular cells.
Polar coordinates are the appropriate
choice for the problem's geometry and should therefore lead to the least
amount of numerical errors. We use periodic and solid-wall boundaries
in the angular and radial directions, respectively. The flow has a maximum Mach
number of $1$ at the inner domain boundary dropping to $\approx 0.6$ at the
outer boundary. We perform most of our simulations with the explicit
RK3 scheme (see \cref{sec:discretization}) because it is more efficient in this Mach number
range. The time step of explicit runs is set by the $\text{CFL}_{uc}$ criterion [\cref{eq:cfluc}]
with $c_\text{CFL}=0.4$. For the lower-resolution setup we also perform simulations with
implicit time stepping and the $\text{CFL}_{ug}$ time step criterion [\cref{eq:cflug}] with $c_\text{CFL}=0.4$.
We find that the results of the implicit runs are identical to the explicit
time stepping (see dots in upper panel of Fig.
\ref{p:keplerian_profiles}). Tests with the explicit, second-order RK2 scheme
also give almost identical results.

Since the disk is isobaric, we use \rhoP reconstruction for this test case.
Here we only compare simulations without \wbing to runs with deviation \wbing,
since the other methods presented in this work are not capable of stabilizing
a target solution with a nonzero velocity field (see \cref{sec:deviation}).

In order to asses the stability of the setup we add a density perturbation
with $\rho = 2$ in the circular region \mbox{$(x+1.5)^2 + y^2
< 0.15^2$} and follow its evolution up to \num{10000} orbital periods, where
the orbital period is taken at the central point of
the perturbation. A perfect solution will maintain the initial radial
density distribution of the perturbation at all times. However, due to its radial
extent there will be a phase shift between the innermost and
outermost regions of the perturbation. Furthermore, numerical
diffusion will spread the perturbation predominantly in the direction of its movement.
Therefore the perturbation evolves into a homogeneous ring orbiting the central
object with a density that corresponds to the initial angle-averaged density
$\overline{\rho}$.

In Fig. \ref{p:keplerian_profiles} we show how the
profile of $\overline{\rho}-1$ evolves with time. The target solution is given
as a black dashed line. For a grid resolution of $20 \times 70$ cells the run
without \wbing only shows small deviations from the target solution after $10$
orbital periods. However, we already see that mass starts to accumulate in
the center. The mass can not leave the domain and accrete onto the central
object due to our wall boundaries. Over time, more and more mass flows to the
inner boundary. After $100$ orbits the center has approximately the same density
as the initial perturbation, and after $1000$ orbits the profile has completely
shifted toward the inner boundary. This is also clearly visible in
the two-dimensional density distribution shown in \cref{fig:keplerian_density}
in \cref{appx:keplerian}. With deviation \wbing the profile remains mostly
stable up to $1000$ orbits. The perturbation is slowly diffusing symmetrically
around the initial peak, maintaining the general shape of the initial density
profile. After $\num{10000}$ orbits we see that also the run with deviation
\wbing has become noticeably asymmetric toward the center. The regions where
the density falls bellow the initial density profile are most likely related to
the fact that we do not use flux limiters for these runs. Undershoots at steep
gradients are a common consequence of this omission.

Increasing the resolution improves the stability of the runs without \wbing
(see bottom panel in Fig. \ref{p:keplerian_profiles}). At a resolution of
$100\times 350$ cells the distribution is still almost identical to the initial
perturbation even after $100$ orbits. Only after $1000$ orbits we start to notice
the accumulation of mass at the inner boundary similar to the low-resolution
case. After $\num{10000}$ orbits, however, the distribution has again completely
shifted towards the center and the initial perturbation is not recognizable any
longer. The high-resolution run with deviation \wbing, on the other hand,
is almost identical to the target solution even after $1000$ orbits. The increased
radial resolution has reduced the radial diffusion observed in the
low-resolution runs. The density undershoots become more noticeable at the
$\num{10000}$ orbit mark. There we again identify a tendency for a slow
drift towards the center. However, thanks to the use of \wbing
the shape of the initial perturbation is still retained approximately.

This test shows that the flexibility of the deviation \wbing method also allows
to maintain stable configurations other than hydrostatic equilibrium. This is
particularly important for the long-term evolution of such systems. Without \wbing
stability can only be achieved by increasing grid resolution,
which leads to substantially higher computational cost.

\section{Summary and conclusions}
\label{sec:conclusions}

We have presented the deviation, the \alphabeta, and the \leroux \wbing methods
that aim to improve the ability of finite-volume codes to maintain hydrostatic
stratifications even at moderate grid resolution. The performance of these
methods were assessed in a set of test simulations of static and dynamical
setups. Special emphasis was given to flows at low Mach numbers. They are
particularly challenging to evolve because they require special low-Mach
hydrodynamic flux solvers, which in turn come with reduced dissipation and
hence are prone to numerical instabilities. Also, it seems natural that slight
deviations from hydrostatic equilibrium lead to low-Mach flows, as is the case,
for example, in stellar convection.  All simulations were performed with the
time-implicit \slh code that solves the fully compressible Euler equations
using a modified version of the low-Mach \ausmpup hydrodynamic flux solver. Our
experience shows that the inclusion of gravitational potential energy in the
total energy is essential to correctly representing slow flows in stratified
atmospheres in the cell-centered discretization of gravity implemented in \slh;
see \citet{mullen2020a} for an alternative approach. To the best of our
knowledge, the present study is the first to reach Mach numbers as low as
\num{2e-4} in stratified convection using the fully compressible Euler
equations.

The first test of the \wbing schemes is to evolve a 1D hydrostatic atmosphere
in time at low resolution for an isothermal, an isentropic, and a polytropic
stratification.  In all cases, the absence of a \wbing scheme quickly led to
spurious velocities at significant amplitudes. The application of any of the
considered \wbing methods removed this problem and managed to keep the flow
below Mach numbers of \num{e-12} for very long times. Repeating the same test
in 2D revealed that low Mach number flux functions, such as \ausmup, are
subject to an exponential growth of the Mach number and horizontal density
fluctuations, which is not physically expected in a stably stratified
atmosphere. This effect became less pronounced the closer the stratification
was to the marginally stable, isentropic profile.  In the isentropic case
\alphabeta and deviation \wbing in combination with the low-Mach flux remained
stable ($\ma \lesssim \num{e-8}$) for long times, while with \leroux \wbing the
setup developed a flow at Mach numbers of about \num{e-1}. These examples
showed that it is important to test \wbd schemes in more than one spatial
dimension and for more than just a few sound-crossing times, as only then
slowly growing instabilities become noticeable, especially in very stable
stratifications.

While it is a necessary condition to maintain an initially static setup, only
dynamical setups are of actual interest in multidimensional simulations. In a
second test, we considered the rise of a hot bubble in a periodic background
stratification spanning two orders of magnitude in pressure at constant entropy.
We tuned the bubble's entropy excess to reach different rising speeds. With
\leroux \wbing and without any \wbing, unphysical entropy fluctuations appeared
in large parts of the atmosphere and a relatively fine grid ($256 \times 384$)
was required to make their amplitude smaller than that of even the hottest
bubble considered. The corresponding Mach number of $\mathrm{Ma} \sim 3 \times
10^{-2}$ seemed to be close to a limit of practical applicability of these two
methods in such a strong stratification. The \alphabeta and deviation methods
fared much better with no entropy changes far from the bubble and only a slight
entropy undershoot right above the bubble even on a coarse ($64 \times 96$)
grid. Equally good results were obtained with the initial amplitude decreased by
a factor of $10^6$, leading to $\mathrm{Ma} \sim 3 \times 10^{-5}$. Numerical
effects started to dominate only at $\mathrm{Ma} \sim 3 \times 10^{-6}$ after
the amplitude was decreased by another factor of $10^2$.

We proceeded with a setup involving a convection zone with a stable zone on
top. The stratification was chosen to be representative of core convection in
a \SI{25}{\msun} main-sequence star. Radiation pressure was a substantial
fraction of the total pressure, testing the methods' capability to deal with
a general \eos. We used volume heating of adjustable amplitude to drive the
convection. With \leroux\wbing and without \wbing, the rms Mach number of the
convective flow ceased to correlate with the heating rate at $\mrms \sim
4 \times 10^{-3}$ and the flow became dominated by small-scale structures of
numerical origin. The lowest reachable value of \mrms dropped by about a factor
of two when switching from \rhoP to \rhoT reconstruction in the absence of
\wbing. Only the \alphabeta and deviation methods were able to reproduce the
expected scaling of \mrms with the heating rate [\cref{eq:mascal}] down to the
slowest flows tested ($\mrms \sim 2 \times 10^{-4}$). We observed spurious
patterns in the stable layer and demonstrated that they were caused by
unresolved internal gravity waves, whose vertical wavelength becomes extremely
short with increasing period. The patterns disappeared when the typical
vertical wavelengths of waves with periods close to the convective turnover
frequency became resolved by $8$ to $10$ cells (grid of $288 \times 2380$
cells).

The deviation method, unlike the other methods presented in this work, can deal
with arbitrary stationary states. We illustrated this capability in our last
test, in which we followed many orbits of a density perturbation in a Keplerian
disk around a point mass. The angle-averaged density profile should ideally
remain constant in this model. Without any \wbing, imperfect balance between
the centrifugal force and gravity led to mass redistribution towards inner
parts of the disk after many orbits. Deviation \wbing much improved the
solutions with only slight radial broadening of the initial perturbation even
after $10^4$ orbits with a moderate resolution of $100 \times 350$ cells.

In summary, our results show that \wbing can substantially reduce the grid resolution
needed to correctly follow tiny perturbations in situations in which the
stationary background state involves the balance of two large opposing forces.
We obtained comparable results with the \alphabeta and deviation methods, both
far surpassing the \leroux method in accuracy in the low-Mach-number regime. The
\alphabeta and deviation methods are also expected to be more accurate than the
majority of other well-balanced methods present in astrophysical literature
\citep[e.g.,][]{zingale2002a,perego2016a,kaeppeli2011a,kaeppeli2016a,padioleau2019a},
because they exactly balance the stationary solution rather than an
approximation to it. Although an analytical prescription is often not
available, the stationary solution can be computed to an arbitrary degree of
accuracy in many astrophysical applications. This statement only holds in the
case that the hydrostatic background state does not change in time. In this
case an update is necessary, which we do not discuss in this paper. This step
would likely involve a local approximation to hydrostatic equilibrium, similar
to the ones suggested in the other aforementioned methods. It should be noted,
however, that such an update is unnecessary in the case of very slow flows,
such as in earlier phases of stellar evolution, because the background state
hardly changes over time, even after hundreds of convective turnover times.

We prefer to use the deviation method, because it is more general and does not
impose any restrictions on reconstruction variables. The method can be applied
both to nearly hydrostatic cases and to cases in which rotation becomes
important. The latter can have strong impact on stellar evolution
\citep{maeder2000a} as well as on the propagation of IGW in stars
\citep{rogers2013a}.

The \leroux method still has its place, though. \citet{horst2021a} show for the
case of convective helium-shell burning that this method considerably reduces
numerical diffusion of the hydrostatic background stratification as compared to
the unbalanced case if the gravitational energy is not included in the total
energy. Therefore we consider the \leroux method as a valid method that leads
to improvements already with little development effort.

\begin{figure*}[h!]
  \centering
  \includegraphics[width=\textwidth]{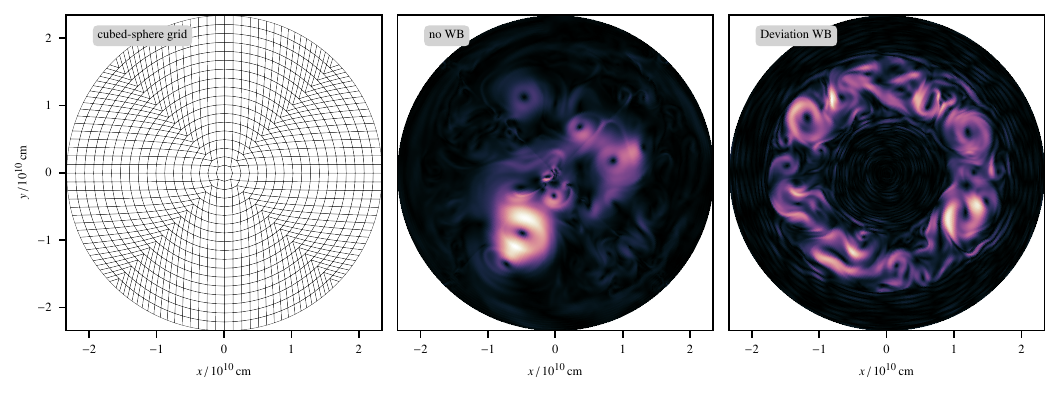}
  \caption{\textbf{Left panel:} Geometry of the cubed-sphere grid. The number
  of cells has been reduced compared to grids used in the other two panels to
  ease the identification of individual cells and their shapes.  \textbf{Middle
  and right panel:} Shell convection in a setup comparable to that in
  \cref{subsec:convection} but with a shallower stratification and ideal gas
  \eos. The middle panel shows the color coded Mach number in a run without
  \wbing. Numerical discretization errors quickly lead to spurious flows in the
  central region and the maximum Mach number reaches \num{3e-2}. In the run
  with deviation \wbing shown in the right panel the convective shell is clearly
  maintained. The maximum Mach number is \num{2e-3}.}

  \label{fig:convbox_pcolor_cubed}

\end{figure*}

This paper focuses on setups with simple grid geometries, but the \slh code can
use general curvilinear grids. This allows us to adapt grid geometry to the
problem at hand. Standard spherical grids adapt to the geometry of slowly or
nonrotating stars, but they suffer from singularities at the center and at
the poles. A star can be inscribed in a simple Cartesian grid
\citep[e.g.,][]{woodward2015a}, but this requires to also impose a
spherical boundary condition on the grid. When done on the cell level, the
sphere is rough on small scales and can generate spurious vorticity. We have
implemented the cubed-sphere grid proposed by \citet{calhoun2008a} in our \slh
code. The grid is logically rectangular but geometrically spherical with a
smooth outer boundary, see \cref{fig:convbox_pcolor_cubed} for a 2D version.
Discretization errors along the grid's strongly deformed diagonals make it
almost impossible to follow nearly hydrostatic flows without any \wbing method.
\Cref{fig:convbox_pcolor_cubed} shows an \slh simulation with a convective shell
embedded between two stably stratified zones. With deviation \wbing, we obtain
the expected convective shell with convection-generated IGW propagating in the
stable zones. If we turn the \wbing off, a convection-like flow of numerical
origin appears and mixes the convection zone with the whole inner stable zone.
This result is promising, but limits of applicability of the deviation method
on the cubed-sphere grid are still to be investigated.

\begin{acknowledgements} PVFE was supported by the U.S. Department of Energy
  through the Los Alamos National Laboratory (LANL). LANL is operated by Triad
  National Security, LLC, for the National Nuclear Security Administration of
  the U.S. Department of Energy (Contract No. 89233218CNA000001). PVFE, LH,
  JPB, RA, JH, GL, and FKR acknowledge support by the Klaus Tschira Foundation.
  The work of FKR and PVFE was supported by the German Research Foundation (DFG)
  through the graduate school on ``Theoretical Astrophysics and Particle
  Physics'' (GRK 1147). The work of CK and FKR is supported by DFG through
  grants KL~566/22-1 and RO~3676/3-1, respectively. JPB acknowledges the grants
  HITS 21.03.2018, HITS 18.12.2018 and HITS 26.08.2020. This work has been assigned
   a document release number LA-UR-21-21056.
\end{acknowledgements}

\bibliographystyle{aa}
\bibliography{./astrofritz}

\begin{thebibliography}{83}
\expandafter\ifx\csname natexlab\endcsname\relax\def\natexlab#1{#1}\fi

\bibitem[{{Almgren} {et~al.}(2006){Almgren}, {Bell}, {Rendleman}, \&
  {Zingale}}]{almgren2006a}
{Almgren}, A.~S., {Bell}, J.~B., {Rendleman}, C.~A., \& {Zingale}, M. 2006,
  \apj, 637, 922

\bibitem[{{Andrassy} {et~al.}(2020){Andrassy}, {Herwig}, {Woodward}, \&
  {Ritter}}]{andrassy2020a}
{Andrassy}, R., {Herwig}, F., {Woodward}, P., \& {Ritter}, C. 2020, \mnras,
  491, 972

\bibitem[{Audusse {et~al.}(2004)Audusse, Bouchut, Bristeau, Klein, \&
  Perthame}]{audusse2004a}
Audusse, E., Bouchut, F., Bristeau, M.-O., Klein, R., \& Perthame, B. 2004,
  SIAM Journal on Scientific Computing, 25, 2050

\bibitem[{Barsukow \& Berberich(2020)}]{barsukow2020b}
Barsukow, W. \& Berberich, J.~P. 2020, Submitted to Journal on Scientific
  Computing

\bibitem[{Barsukow {et~al.}(2017)Barsukow, Edelmann, Klingenberg, Miczek, \&
  R{\"o}pke}]{barsukow2017a}
Barsukow, W., Edelmann, P. V.~F., Klingenberg, C., Miczek, F., \& R{\"o}pke,
  F.~K. 2017, Journal of Scientific Computing, 72, 623

\bibitem[{{Barsukow} {et~al.}(2017){Barsukow}, {Edelmann}, {Klingenberg}, \&
  {R{\"o}pke}}]{barsukow2017b}
{Barsukow}, W., {Edelmann}, P. V.~F., {Klingenberg}, C., \& {R{\"o}pke}, F.~K.
  2017, in ESAIM: Proceedings and Surveys, Vol.~58, Workshop on low velocity
  flows, Paris, 5--6 November 2015, ed. S.~Dellacherie \& {et al.}, 27--39

\bibitem[{Berberich(2021)}]{berberich2021b}
Berberich, J.~P. 2021, doctoralthesis, Universit{\"a}t W{\"u}rzburg

\bibitem[{Berberich {et~al.}(2018)Berberich, Chandrashekar, \&
  Klingenberg}]{berberich2018a}
Berberich, J.~P., Chandrashekar, P., \& Klingenberg, C. 2018, in Theory,
  Numerics and Applications of Hyperbolic Problems I, Springer Proceedings in
  Mathematics \& Statistics 236, ed. {C.\ Klingenberg} \& {M.\ Westdickenberg}

\bibitem[{Berberich {et~al.}(2021)Berberich, Chandrashekar, \&
  Klingenberg}]{berberich2021a}
Berberich, J.~P., Chandrashekar, P., \& Klingenberg, C. 2021, Computers \&
  Fluids, 104858

\bibitem[{Berberich {et~al.}(2019)Berberich, Chandrashekar, Klingenberg, \&
  R\"opke}]{berberich2019a}
Berberich, J.~P., Chandrashekar, P., Klingenberg, C., \& R\"opke, F.~K. 2019,
  Communications in Computational Physics

\bibitem[{{Berberich} {et~al.}(2020){Berberich}, {K{\"a}ppeli},
  {Chandrashekar}, \& {Klingenberg}}]{berberich2020c}
{Berberich}, J.~P., {K{\"a}ppeli}, R., {Chandrashekar}, P., \& {Klingenberg},
  C. 2020, Accepted in Communications in Computational Physics,
  arXiv:2005.01811

\bibitem[{Berberich \& Klingenberg(2020)}]{berberich2020b}
Berberich, J.~P. \& Klingenberg, C. 2020, Accepted for publication in: SEMA
  SIMAI Series: Numerical methods for hyperbolic problems Numhyp 2019

\bibitem[{Bermudez \& V{\'a}zquez(1994)}]{bermudez1994a}
Bermudez, A. \& V{\'a}zquez, M.~E. 1994, Computers \& Fluids, 23, 1049

\bibitem[{{Bola{\~n}os Rosales}(2016)}]{bolanos_phd}
{Bola{\~n}os Rosales}, A. 2016, Dissertation, Julius-Maximilians-Universit\"at
  W\"urzburg

\bibitem[{{Browning} {et~al.}(2004){Browning}, {Brun}, \&
  {Toomre}}]{browning2004a}
{Browning}, M.~K., {Brun}, A.~S., \& {Toomre}, J. 2004, \apj, 601, 512

\bibitem[{Brufau {et~al.}(2002)Brufau, V{\'a}zquez-Cend{\'o}n, \&
  Garc{\'\i}a-Navarro}]{brufau2002a}
Brufau, P., V{\'a}zquez-Cend{\'o}n, M., \& Garc{\'\i}a-Navarro, P. 2002,
  International Journal for Numerical Methods in Fluids, 39, 247

\bibitem[{{Calhoun} {et~al.}(2008){Calhoun}, {Helzel}, \&
  {Leveque}}]{calhoun2008a}
{Calhoun}, D.~A., {Helzel}, C., \& {Leveque}, R.~J. 2008, SIAM Review, 50, 723

\bibitem[{Cargo \& Le~Roux(1994)}]{cargo1994a}
Cargo, P. \& Le~Roux, A. 1994, Comptes rendus de l'Acad{\'e}mie des sciences.
  S{\'e}rie 1, Math{\'e}matique, 318, 73

\bibitem[{Castro \& Semplice(2018)}]{castro2018a}
Castro, M.~J. \& Semplice, M. 2018, International Journal for Numerical Methods
  in Fluids

\bibitem[{Chandrashekar \& Klingenberg(2015)}]{chandrashekar2015a}
Chandrashekar, P. \& Klingenberg, C. 2015, SIAM Journal on Scientific
  Computing, 37, B382

\bibitem[{{Courant} {et~al.}(1928){Courant}, {Friedrichs}, \&
  {Lewy}}]{courant1928a}
{Courant}, R., {Friedrichs}, K.~O., \& {Lewy}, H. 1928, Math. Ann., 100, 32

\bibitem[{{Cristini} {et~al.}(2019){Cristini}, {Hirschi}, {Meakin}, {Arnett},
  {Georgy}, \& {Walkington}}]{cristini2019a}
{Cristini}, A., {Hirschi}, R., {Meakin}, C., {et~al.} 2019, \mnras, 484, 4645

\bibitem[{{Cristini} {et~al.}(2017){Cristini}, {Meakin}, {Hirschi}, {Arnett},
  {Georgy}, {Viallet}, \& {Walkington}}]{cristini2017a}
{Cristini}, A., {Meakin}, C., {Hirschi}, R., {et~al.} 2017, \mnras, 471, 279

\bibitem[{{Dedner} {et~al.}(2001){Dedner}, {Kr{\"o}ner}, {Sofronov}, \&
  {Wesenberg}}]{dedner2001a}
{Dedner}, A., {Kr{\"o}ner}, D., {Sofronov}, I.~L., \& {Wesenberg}, M. 2001,
  Journal of Computational Physics, 171, 448

\bibitem[{Desveaux {et~al.}(2016{\natexlab{a}})Desveaux, Zenk, Berthon, \&
  Klingenberg}]{desveaux2016a}
Desveaux, V., Zenk, M., Berthon, C., \& Klingenberg, C. 2016{\natexlab{a}},
  International Journal for Numerical Methods in Fluids, 81, 104, fld.4177

\bibitem[{Desveaux {et~al.}(2016{\natexlab{b}})Desveaux, Zenk, Berthon, \&
  Klingenberg}]{desveaux2016b}
Desveaux, V., Zenk, M., Berthon, C., \& Klingenberg, C. 2016{\natexlab{b}},
  Mathematics of Computation, 85, 1571

\bibitem[{{Edelmann}(2014)}]{edelmann2014a}
{Edelmann}, P.~V.~F. 2014, Dissertation, Technische Universit\"at M\"unchen

\bibitem[{{Edelmann} {et~al.}(2019){Edelmann}, {Ratnasingam}, {Pedersen},
  {Bowman}, {Prat}, \& {Rogers}}]{edelmann2019a}
{Edelmann}, P.~V.~F., {Ratnasingam}, R.~P., {Pedersen}, M.~G., {et~al.} 2019,
  \apj, 876, 4

\bibitem[{{Edelmann} \& {R\"{o}pke}(2016)}]{edelmann2016b}
{Edelmann}, P.~V.~F. \& {R\"{o}pke}, F.~K. 2016, in {JUQUEEN} {E}xtreme
  {S}caling {W}orkshop 2016, ed. D.~Br\"ommel, W.~Frings, \& B.~J.~N. Wylie,
  JSC Internal Report No. FZJ-JSC-IB-2016-01, 63--67

\bibitem[{{Edelmann} {et~al.}(2017){Edelmann}, {R{\"o}pke}, {Hirschi},
  {Georgy}, \& {Jones}}]{edelmann2017a}
{Edelmann}, P.~V.~F., {R{\"o}pke}, F.~K., {Hirschi}, R., {Georgy}, C., \&
  {Jones}, S. 2017, \aap, 604, A25

\bibitem[{Edwards \& Liou(1998)}]{edwards1998a}
Edwards, J.~R. \& Liou, M.-S. 1998, AIAA journal, 36, 1610

\bibitem[{{Gaburro} {et~al.}(2018){Gaburro}, {Castro}, \&
  {Dumbser}}]{gaburro2018a}
{Gaburro}, E., {Castro}, M.~J., \& {Dumbser}, M. 2018, \mnras, 477, 2251

\bibitem[{{Goffrey} {et~al.}(2017){Goffrey}, {Pratt}, {Viallet}, {Baraffe},
  {Popov}, {Walder}, {Folini}, {Geroux}, \& {Constantino}}]{goffrey2017a}
{Goffrey}, T., {Pratt}, J., {Viallet}, M., {et~al.} 2017, \aap, 600, A7

\bibitem[{Grosheintz-Laval \& K{\"a}ppeli(2019)}]{grosheintz2019a}
Grosheintz-Laval, L. \& K{\"a}ppeli, R. 2019, Journal of Computational Physics,
  378, 324

\bibitem[{{Guillard} \& {Murrone}(2004)}]{guillard2004a}
{Guillard}, H. \& {Murrone}, A. 2004, Computers \& Fluids, 33, 655

\bibitem[{Guillard \& Viozat(1999)}]{guillard1999a}
Guillard, H. \& Viozat, C. 1999, Computers \& Fluids, 28, 63

\bibitem[{{Higl, J.} {et~al.}(2021){Higl, J.}, {M\"uller, E.}, \& {Weiss,
  A.}}]{higl2020a}
{Higl, J.}, {M\"uller, E.}, \& {Weiss, A.} 2021, \aap, 646, A133

\bibitem[{{Horst} {et~al.}(2020){Horst}, {Edelmann}, {Andr{\'a}ssy},
  {R{\"o}pke}, {Bowman}, {Aerts}, \& {Ratnasingam}}]{horst2020a}
{Horst}, L., {Edelmann}, P.~V.~F., {Andr{\'a}ssy}, R., {et~al.} 2020, \aap,
  641, A18

\bibitem[{{Horst} {et~al.}(2021){Horst}, {Hirschi}, {Edelmann}, {Andrassy}, \&
  {Roepke}}]{horst2021a}
{Horst}, L., {Hirschi}, R., {Edelmann}, P.~V.~F., {Andrassy}, R., \& {Roepke},
  F.~K. 2021, arXiv e-prints, arXiv:2107.02199

\bibitem[{{Hosea} \& {Shampine}(1996)}]{hosea1996a}
{Hosea}, M. \& {Shampine}, L. 1996, Applied Numerical Mathematics, 20, 21 ,
  method of Lines for Time-Dependent Problems

\bibitem[{{Jones} {et~al.}(2017){Jones}, {Andrassy}, {Sandalski}, {Davis},
  {Woodward}, \& {Herwig}}]{jones2017a}
{Jones}, S., {Andrassy}, R., {Sandalski}, S., {et~al.} 2017, \mnras, 465, 2991

\bibitem[{K{\"a}ppeli \& Mishra(2014)}]{kaeppeli2014a}
K{\"a}ppeli, R. \& Mishra, S. 2014, Journal of Computational Physics, 259, 199

\bibitem[{K{\"a}ppeli \& Mishra(2016)}]{kaeppeli2016a}
K{\"a}ppeli, R. \& Mishra, S. 2016, Astronomy \& Astrophysics, 587, A94

\bibitem[{{K{\"a}ppeli} {et~al.}(2011){K{\"a}ppeli}, {Whitehouse},
  {Scheidegger}, {Pen}, \& {Liebend{\"o}rfer}}]{kaeppeli2011a}
{K{\"a}ppeli}, R., {Whitehouse}, S.~C., {Scheidegger}, S., {Pen}, U.~L., \&
  {Liebend{\"o}rfer}, M. 2011, \apjs, 195, 20

\bibitem[{{Kifonidis} \& {M{\"u}ller}(2012)}]{kifonidis2012a}
{Kifonidis}, K. \& {M{\"u}ller}, E. 2012, \aap, 544, A47

\bibitem[{{Kippenhahn} {et~al.}(2012){Kippenhahn}, {Weigert}, \&
  {Weiss}}]{kippenhahn2012a}
{Kippenhahn}, R., {Weigert}, A., \& {Weiss}, A. 2012, {Stellar Structure and
  Evolution} (Berlin Heidelberg: Springer-Verlag)

\bibitem[{{Le Roux}(1999)}]{leroux1999a}
{Le Roux}, A.~Y. 1999, ESAIM: Proc., 6, 75

\bibitem[{LeVeque(1998)}]{leveque1998b}
LeVeque, R.~J. 1998, Journal of computational physics, 146, 346

\bibitem[{Li \& Gu(2008)}]{li2008a}
Li, X.-s. \& Gu, C.-w. 2008, Journal of Computational Physics, 227, 5144

\bibitem[{Liou(1996)}]{liou1996a}
Liou, M.-S. 1996, Journal of Computational Physics, 129, 364

\bibitem[{Liou(2006)}]{liou2006a}
Liou, M.-S. 2006, Journal of Computational Physics, 214, 137

\bibitem[{Liou \& Steffen(1993)}]{liou1993a}
Liou, M.-S. \& Steffen, C. J.~J. 1993, Journal of Computational Physics, 107,
  23

\bibitem[{{Maeder}(2009)}]{maeder2009a}
{Maeder}, A. 2009, {Physics, Formation and Evolution of Rotating Stars},
  Astronomy and Astrophysics Library (Springer Berlin Heidelberg)

\bibitem[{{Maeder} \& {Meynet}(2000)}]{maeder2000a}
{Maeder}, A. \& {Meynet}, G. 2000, \araa, 38, 143

\bibitem[{{Meakin} \& {Arnett}(2006)}]{meakin2006a}
{Meakin}, C.~A. \& {Arnett}, D. 2006, \apjl, 637, L53

\bibitem[{{Meakin} \& {Arnett}(2007)}]{meakin2007a}
{Meakin}, C.~A. \& {Arnett}, D. 2007, \apj, 667, 448

\bibitem[{{Michel}(2019)}]{michel_phd}
{Michel}, A. 2019, Dissertation, Ruprecht-Karls-Universit\"at Heidelberg

\bibitem[{Miczek(2013)}]{miczek2013a}
Miczek, F. 2013, Dissertation, Technische Universit\"at M\"unchen

\bibitem[{{Miczek} {et~al.}(2015){Miczek}, {R{\"o}pke}, \&
  {Edelmann}}]{miczek2015a}
{Miczek}, F., {R{\"o}pke}, F.~K., \& {Edelmann}, P.~V.~F. 2015, \aap, 576, A50

\bibitem[{{Mullen} {et~al.}(2020){Mullen}, {Hanawa}, \& {Gammie}}]{mullen2020a}
{Mullen}, P.~D., {Hanawa}, T., \& {Gammie}, C.~F. 2020, arXiv e-prints,
  arXiv:2012.01340

\bibitem[{{M{\"u}ller} {et~al.}(2016){M{\"u}ller}, {Viallet}, {Heger}, \&
  {Janka}}]{mueller2016a}
{M{\"u}ller}, B., {Viallet}, M., {Heger}, A., \& {Janka}, H.-T. 2016, \apj,
  833, 124

\bibitem[{O{\ss}wald {et~al.}(2015)O{\ss}wald, Siegmund, Birken, Hannemann, \&
  Meister}]{osswald2015a}
O{\ss}wald, K., Siegmund, A., Birken, P., Hannemann, V., \& Meister, A. 2015,
  International Journal for Numerical Methods in Fluids, fld.4175

\bibitem[{Padioleau {et~al.}(2019)Padioleau, Tremblin, Audit, Kestener, \&
  Kokh}]{padioleau2019a}
Padioleau, T., Tremblin, P., Audit, E., Kestener, P., \& Kokh, S. 2019, The
  Astrophysical Journal, 875, 128

\bibitem[{{Perego} {et~al.}(2016){Perego}, {Cabez{\'o}n}, \&
  {K{\"a}ppeli}}]{perego2016a}
{Perego}, A., {Cabez{\'o}n}, R.~M., \& {K{\"a}ppeli}, R. 2016, Astrophysical
  Journal, Supplement, 223, 22

\bibitem[{{Popov} {et~al.}(2019){Popov}, {Walder}, {Folini}, {Goffrey},
  {Baraffe}, {Constantino}, {Geroux}, {Pratt}, {Viallet}, \&
  {K{\"a}ppeli}}]{popov2019a}
{Popov}, M.~V., {Walder}, R., {Folini}, D., {et~al.} 2019, \aap, 630, A129

\bibitem[{{Pratt} {et~al.}(2020){Pratt}, {Baraffe}, {Goffrey}, {Geroux},
  {Constantino}, {Folini}, \& {Walder}}]{pratt2020a}
{Pratt}, J., {Baraffe}, I., {Goffrey}, T., {et~al.} 2020, \aap, 638, A15

\bibitem[{{Pratt} {et~al.}(2016){Pratt}, {Baraffe}, {Goffrey}, {Geroux},
  {Viallet}, {Folini}, {Constantino}, {Popov}, \& {Walder}}]{pratt2016a}
{Pratt}, J., {Baraffe}, I., {Goffrey}, T., {et~al.} 2016, \aap, 593, A121

\bibitem[{Rieper(2011)}]{rieper2011a}
Rieper, F. 2011, Journal of Computational Physics, 230, 5263

\bibitem[{Roe(1981)}]{roe1981a}
Roe, P.~L. 1981, Journal of Computational Physics, 43, 357

\bibitem[{{Rogers} {et~al.}(2013){Rogers}, {Lin}, {McElwaine}, \&
  {Lau}}]{rogers2013a}
{Rogers}, T.~M., {Lin}, D.~N.~C., {McElwaine}, J.~N., \& {Lau}, H.~H.~B. 2013,
  \apj, 772, 21

\bibitem[{R{\"o}pke {et~al.}(2018)R{\"o}pke, Berberich, Edelmann, Horst, Jones,
  Michel, \& Botto~Poala}]{roepke2018b}
R{\"o}pke, F.~K., Berberich, J., Edelmann, P. F.~V., {et~al.} 2018, in NIC
  Series, Vol.~49, NIC Symposium 2018, NIC Symposium 2018, J{\"u}lich
  (Germany), 22 Feb 2018 - 23 Feb 2018 (J{\"u}lich: Forschungszentrum
  J{\"u}lich GmbH, Zentralbibliothek), 115 -- 122

\bibitem[{{Shu} \& {Osher}(1988)}]{shu1988a}
{Shu}, C.-W. \& {Osher}, S. 1988, Journal of Computational Physics, 77, 439

\bibitem[{Sutherland(2010)}]{sutherland2010a}
Sutherland, B. 2010, Internal Gravity Waves (Cambridge University Press)

\bibitem[{{Tassoul}(2000)}]{tassoul2000a}
{Tassoul}, J.-L. 2000, {Stellar Rotation}

\bibitem[{{Timmes} \& {Swesty}(2000)}]{timmes2000a}
{Timmes}, F.~X. \& {Swesty}, F.~D. 2000, \apjs, 126, 501

\bibitem[{Toro(2009)}]{toro2009a}
Toro, E.~F. 2009, Riemann Solvers and Numerical Methods for Fluid Dynamics: A
  Practical Introduction (Berlin Heidelberg: Springer)

\bibitem[{Touma \& Klingenberg(2015)}]{touma2015a}
Touma, R. \& Klingenberg, C. 2015, Applied Numerical Mathematics, 97, 42

\bibitem[{Touma {et~al.}(2016)Touma, Koley, \& Klingenberg}]{touma2016a}
Touma, R., Koley, U., \& Klingenberg, C. 2016, SIAM Journal on Scientific
  Computing, 38, B773

\bibitem[{Turkel(1987)}]{turkel1987a}
Turkel, E. 1987, Journal of computational physics, 72, 277

\bibitem[{{Veiga} {et~al.}(2019){Veiga}, {Romero Velasco}, {Abgrall}, \&
  {Teyssier}}]{veiga2019a}
{Veiga}, M.~H., {Romero Velasco}, D.~A., {Abgrall}, R., \& {Teyssier}, R. 2019,
  Communications in Computational Physics, 26, 1

\bibitem[{{Viallet} {et~al.}(2013){Viallet}, {Meakin}, {Arnett}, \&
  {Moc{\'a}k}}]{viallet2013b}
{Viallet}, M., {Meakin}, C., {Arnett}, D., \& {Moc{\'a}k}, M. 2013, \apj, 769,
  1

\bibitem[{{Woodward} {et~al.}(2015){Woodward}, {Herwig}, \&
  {Lin}}]{woodward2015a}
{Woodward}, P.~R., {Herwig}, F., \& {Lin}, P.-H. 2015, \apj, 798, 49

\bibitem[{{Zingale} {et~al.}(2002){Zingale}, {Dursi}, {ZuHone}, {Calder},
  {Fryxell}, {Plewa}, {Truran}, {Caceres}, {Olson}, {Ricker}, {Riley},
  {Rosner}, {Siegel}, {Timmes}, \& {Vladimirova}}]{zingale2002a}
{Zingale}, M., {Dursi}, L.~J., {ZuHone}, J., {et~al.} 2002, \apjs, 143, 539

\end{thebibliography}

\begin{appendix}
\section{Simple stratified atmospheres}
\label{appx:hystattest}
These are the 1D counterparts of the isentropic and polytropic tests in \cref{sec:hystattest}.
\begin{figure}
  \includegraphics[width=\columnwidth]{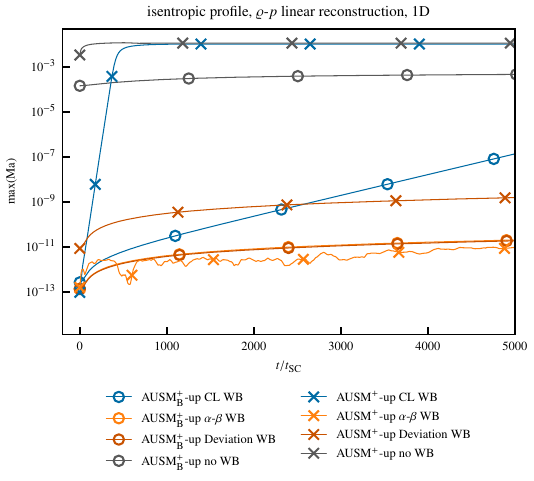}
  \caption{\label{fig:hystattest-isentropic-1d}Same as
    \cref{fig:hystattest-isentropic-2d}, but as a 1D simulation. The adiabatic
    exponent is $\gamma=5/3$. The solid lines represent the maximum Mach
    numbers on the grid. Time is given in units of the sound-crossing
    time~$\tSC=\SI{4.28}{s}$. The curves have been slightly smoothed for better
  visibility.}
\end{figure}

\begin{figure}
  \includegraphics[width=\columnwidth]{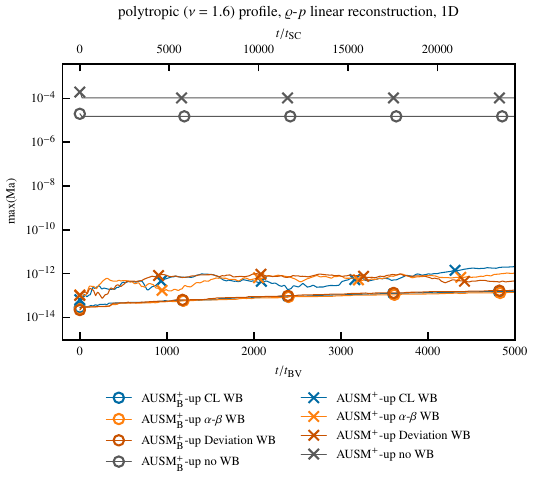}
  \caption{\label{fig:hystattest-polytropic-nu-1_6-1d}Same as
  \cref{fig:hystattest-polytropic-nu-1_6-2d}, but as a 1D simulation. The
  adiabatic exponent is $\gamma=5/3$. The solid lines represent the maximum
  Mach numbers on the grid. Time is given in units of \brunt
  time~$\tBV=\SI{20.1}{s}$ and sound-crossing time~$\tSC=\SI{4.13}{s}$. The
  curves have been slightly smoothed for better visibility.}
\end{figure}

\section{Hot bubble}
\label{appx:hotbubble}
This appendix explores the amplitude dependence of the hot bubble test from \cref{sec:hotbubble}.
\begin{figure*}
  \centering
  \includegraphics[width=0.95\textwidth]{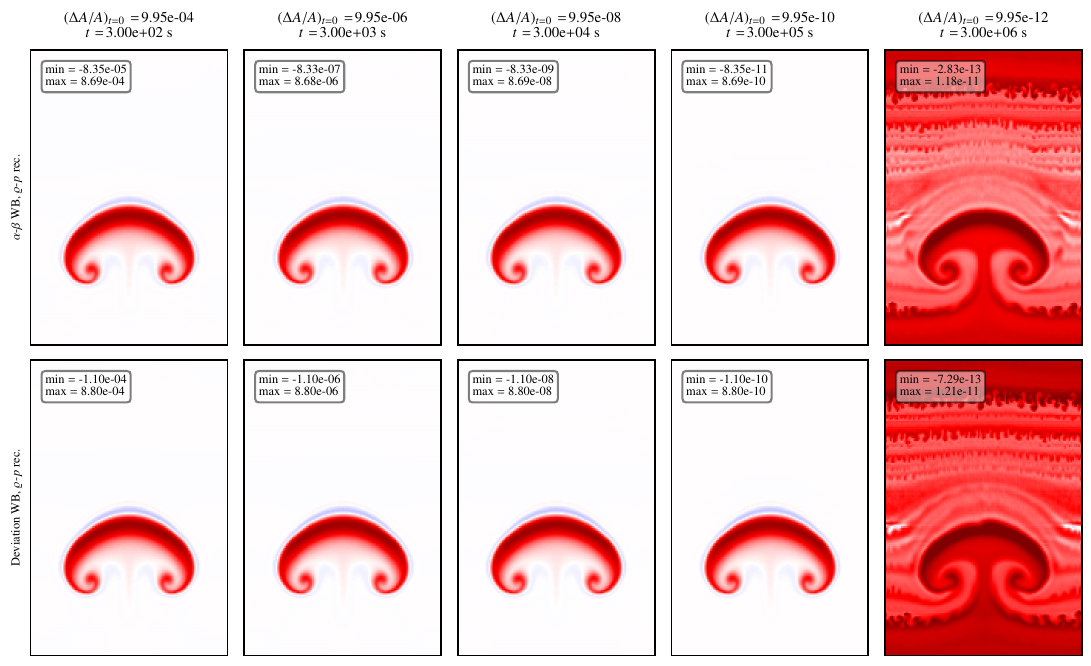}
  \caption{Same as \cref{fig:hotbubble-time-dependence}, but showing the
      dependence of the solution on the initial amplitude $(\Delta A/A)_{t=0}$
      (left to right) on the $128 \times 192$ grid for the two best
      well-balancing methods (rows). The final time of the simulations is
      scaled with $(\Delta A/A)_{t=0}^{1/2}$ allowing the bubbles to reach the
      same evolutionary stage, see \cref{sec:hotbubble}.}
  \label{fig:hotbubble-amplitude-dependence}
\end{figure*}

\begin{figure*}
  \centering
  \includegraphics[width=0.95\textwidth]{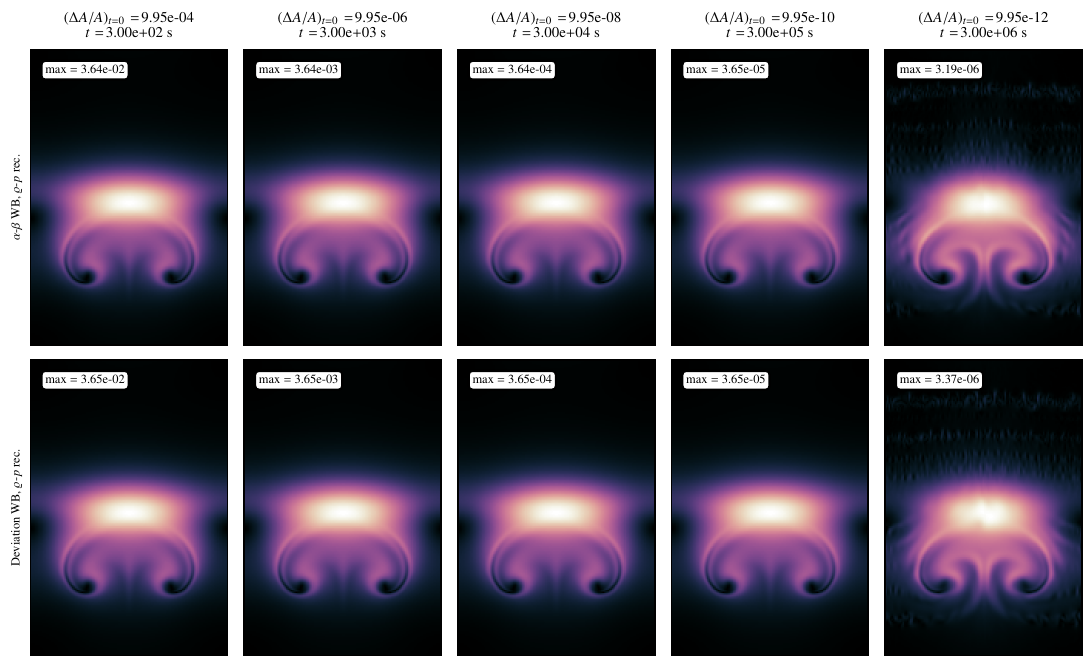}
  \caption{Mach-number distributions in the solutions shown in
  Fig.~\ref{fig:hotbubble-amplitude-dependence}. The color scheme ranges from
  $0$ (black) to the maximum value (white) reached in the simulation box, which
  is also indicated at the top of the panels.}
  \label{fig:hotbubble-amplitude-dependence-Ma}
\end{figure*}

\section{Simple convective box setup}
\label{appx:convbox}
\Cref{tab:hsedev} shows that the relative deviation from hydrostatic equilibrium in the convective box test from \cref{subsec:convection} scales with the square of the rms Mach number in the convection zone.
\begin{table}
  \caption{Relative deviation from hydrostatic equilibrium, rms Mach number,
  and ratio of the squared Mach number to the deviation at different
  heating rates.}
  \label{tab:hsedev}
  \begin{tabular}{c|c|c|c|c|}
  $\dote$ & $\frac{(\partial p/\partial y) - \rho\,g}{(\partial p/\partial y)}$ & $\mrms$ & $\mrms^2\left[\frac{(\partial p/\partial y)-\rho\,g}{(\partial p/\partial y)}\right]^{-1}$\\
  \hline\rule{0pt}{2.6ex}\rule[-1.2ex]{0pt}{0pt}
    \num{e+02}& \num{2.03e-07}& \num{2.07e-04}& \num{0.21} \\
    \num{e+03}& \num{1.10e-06}& \num{4.54e-04}& \num{0.19} \\
    \num{e+04}& \num{3.71e-06}& \num{9.50e-04}& \num{0.24} \\
    \num{e+05}& \num{1.45e-05}& \num{2.04e-03}& \num{0.29} \\
    \num{e+06}& \num{3.79e-05}& \num{4.33e-03}& \num{0.49} \\
    \num{e+07}& \num{1.88e-04}& \num{9.21e-03}& \num{0.45} \\
  \end{tabular}
  \tablefoot{The data is derived from the simulations using the \alphabeta
    \wbing method, spatially averaged over the convection zone, and averaged
    over a time frame spanning $5\,\tconv$. The results confirm the expected
    scaling of the relative deviation from hydrostatic equilibrium with
    $\sim\mrms^2$ (see text) within the accuracy of our order-of-magnitude
    estimate.}
\end{table}

\begin{figure}
  \includegraphics[width=\columnwidth]{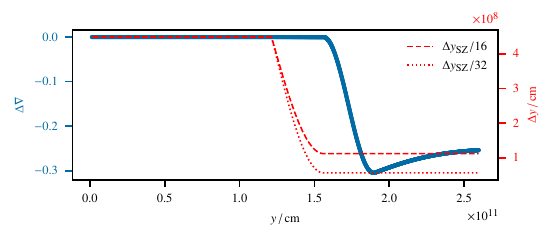}
  \caption{Varying vertical grid spacing as a function of the vertical coordinate $y$
  for the simulations shown in the center and right panels of the lower row
  in \cref{fig:convbox_pcolor_pat}. The superadiabaticity is shown as a blue
  line, a negative value indicates a convectively stable stratification. The
  spacing changes smoothly to a finer resolution slightly before the
  transition to the stable zone starts.}
  \label{fig:convbox_vary}
\end{figure}

\section{Keplerian disk}
\label{appx:keplerian}
\begin{figure*}
  \centering
  \includegraphics[width=\textwidth]{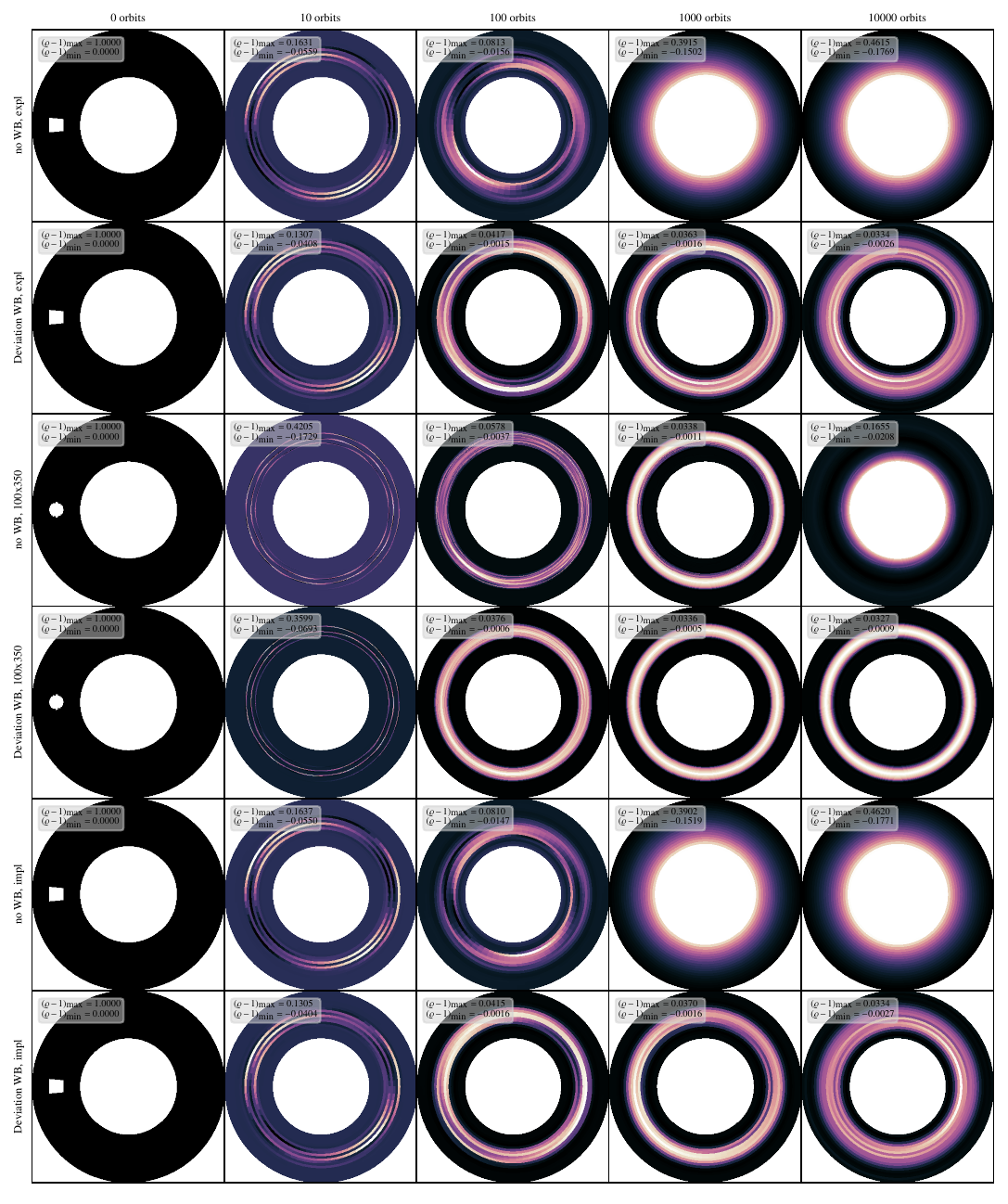}
  \caption{Snapshots of the density
     distribution during the evolution of the Keplerian disk. Shown are runs
     with deviation \wbing and without \wbing using explicit time stepping and
     resolutions of $20 \times 70$ cells as well as $100 \times 350$ grid cells.
     The bottom two rows show the same setup evolved with implicit time
     stepping at a resolution of $20 \times 70$ cells. To emphasize the
     deviation from the initial background density of $\rho_0 = 1$ we show here
  $\rho - 1$ and give the maximum and minimum of this quantity in an inset for
  each snapshot.}
  \label{fig:keplerian_density}
\end{figure*}

\Cref{fig:keplerian_density} shows the time evolution of density in the Keplerian disk problem from \cref{sec:keplerian}.

\end{appendix}
\end{document}